\documentclass[journal,comsoc]{IEEEtran}
\usepackage[T1]{fontenc}
\usepackage{cite}
\usepackage{algorithmic}
\usepackage{graphicx}
\usepackage{textcomp}
\usepackage{xcolor}
\usepackage{url}
\usepackage{tabu}
\usepackage{multirow}
\usepackage[normalem]{ulem}
\useunder{\uline}{\ul}{}
\usepackage{makecell} 
\usepackage{color}

\ifCLASSINFOpdf
\else
\fi
\usepackage{amsmath}
\usepackage[cmintegrals]{newtxmath}
\begin{document}
%


\title{Smart Grid: Cyber Attacks, Critical Defense Approaches, and Digital Twin}

%

\author{Tianming~Zheng,
        Ming~Liu,
        Deepak~Puthal,
        Ping~Yi*,~\IEEEmembership{Senior~Member,~IEEE,}
        Yue~Wu*,~\IEEEmembership{Member,~IEEE,}
        and~Xiangjian~He,~\IEEEmembership{Senior~Member,~IEEE}
\thanks{This work was supported by the National Key R\&D Program of China under Grants No. 2020YFB1807500, No. 2020YFB1807504, and National Science Foundation of China Key Project under Grants No. 61831007. \textit{(Corresponding authors: Ping Yi, Yue Wu.)}}
\thanks{T. Zheng, M. Liu, P. Yi, and Y. Wu are with the School of Electronic Information and Electrical Engineering, Shanghai Jiao Tong University, Shanghai, 200240 China (e-mail: zhengtianming@sjtu.edu.cn; liuming198904@sjtu.edu.cn; yiping@sjtu.edu.cn; wuyue@sjtu.edu.cn).}
\thanks{D. Puthal is with the School of Computing, Newcastle University, UK (e-mail: deepak.puthal@gmail.com).}
\thanks{X. He is with the School of Electrical and Data Engineering, University of Technology Sydney, Australia (e-mail: Xiangjian.He@uts.edu.au).}
}

\maketitle

\begin{abstract}

As a national critical infrastructure, the smart grid has attracted widespread attention for its cybersecurity issues. The development towards an intelligent, digital, and Internet-connected smart grid has attracted external adversaries for malicious activities. It is necessary to enhance its cybersecurity by either improving the existing defense approaches or introducing novel developed technologies to the smart grid context. As an emerging technology, digital twin (DT) is considered as an enabler for enhanced security. However, the practical implementation is quite challenging.
This is due to the knowledge barriers among smart grid designers, security experts, and DT developers. Each single domain is a complicated system covering various components and technologies.
As a result, works are needed to sort out relevant contents so that DT can be better embedded in the security architecture design of smart grid.

In order to meet this demand, our paper covers the above three domains, i.e., smart grid, cybersecurity, and DT.
Specifically, the paper i) introduces the background of the smart grid;
ii) reviews external cyber attacks from attack incidents and attack methods;
iii) introduces critical defense approaches in industrial cyber systems, which include device identification, vulnerability discovery, intrusion detection systems (IDSs), honeypots, attribution, and threat intelligence (TI);
iv) reviews the relevant content of DT, including its basic concepts, applications in the smart grid, and how DT enhances the security.
In the end, the paper puts forward our security considerations on the future development of DT-based smart grid.
The survey is expected to help developers break knowledge barriers among smart grid, cybersecurity, and DT, and provide guidelines for future security design of DT-based smart grid.

\end{abstract}


\section{Introduction}

\IEEEPARstart{A}{s} a national critical infrastructure, the smart grid has attracted widespread attention from governments, industries, and academia. 
A market research \cite{Market2018Market} predicted that smart grid's market would increase from USD 23.8 billion to USD 61.3 billion from 2018 to 2023. 
However, the bright future is often accompanied by challenges. The smart grid developing towards an intelligent, digital, and Internet-connected cyber-physical system (CPS) has also expanded the threat surface, which attracted external adversaries for malicious activities.
To fully understand its cybersecurity issues, a review of external cyberattacks, including last-decade cyberattack incidents and attack methods, can be helpful.

Besides, to ensure the trustworthiness of the smart grid, security measures need to be updated for enhanced safety, reliability, security, resilience, and privacy \cite{CPS2017NIST}. 
To achieve the goal, efforts can be made by either improving the existing defense approaches or applying novel developed technologies in the smart grid context.

Existing security measures like device identification, vulnerability discovery, intrusion detection system (IDS), honeypot, attribution, and threat intelligence (TI) have been correlated to form a systematic passive-active defense architecture.
They could be used to identify suspicious devices running vulnerable software, discover malicious host and network behaviors, distract adversaries' attention with deliberately deployed devices, track adversary's identity, and generate reports to guide the enforcement of smart grid's security.
Besides, with the development of artificial intelligence (AI), most technologies have been improved with AI algorithms. It has significantly reduced security analyst's work and improved the performance of defense approaches.

Moreover, digital twin (DT) has become a promising technology in various industry and network scenarios. 
It acts as a virtual representation of the real-world entity or system \cite{GartnerDTdefine}. 
It is initially proposed by Grieves in 2003 for product manufacturing process \cite{grieves2014digital}. Until recently, its development has received extensive attention.
In the technology trends for 2021, Accenture regarded DT as the top five strategy technologies \cite{Techvision2021}. A market analysis indicated that the global DT market size was expected to increase from USD 3.1 billion to USD 48.2 billion from 2020 to 2026 \cite{DTmarket}. 

\begin{figure}[htbp]
\centerline{\includegraphics[scale=0.7]{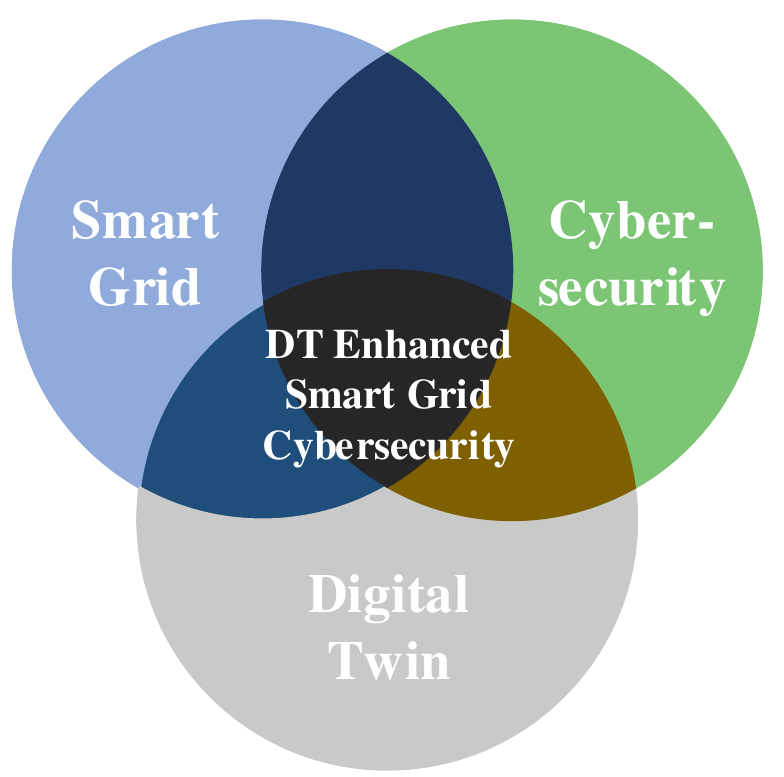}}
\caption{Scope of the survey.}
\label{Figure: three domain}
\end{figure}

It is worth noticing that DT is regarded as an enabler for enhanced security\cite{gehrmann2019digital,bitton2018deriving,eckhart2018towards,eckhart2018specification,gehrmann2016iot,ITEACyberFactory,becue2018cyberfactory}.
However, many existing works still focus on analyzing the concept itself, figuring out its components, or discussing the DT framework. Instead, only a very limited amount of work has actually been practiced in the realistic smart grid-cybersecurity context. 
The reasons are in two aspects.
  Firstly, the technology is still in a very early stage. Many researchers haven't clearly understand the meaning and effect of DT, not to mention its technical details or concrete applications. 
  Secondly, a secure DT-based system is quite complicated. It requires the knowledge of various domains, including SMART GRID, CYBERSECURITY, and DT itself. The knowledge barriers make the study of the interdisciplinary work quite challenging.
Therefore, there is a urgent need to break down the knowledge barriers, sort out related content, introduce required technologies, and provide guidelines for DT enhanced industrial cybersecurity.

In order to meet these demands, our survey reviews the relevant content covering the three domains, i.e., "SMART GRID, CYBERSECURITY, and DT". 
Specifically, the paper reviews smart grid cyberattacks, critical defense approaches, and existing works of DT.
The survey is expected to help developers break knowledge barriers among smart grid, cybersecurity, and DT, and provide guidelines for future security design of DT-based smart grid.
The scope of the survey is illustrated in Fig.~\ref{Figure: three domain}.
The contributions are listed as follows:

\begin{itemize} 

  \item We have reviewed and analyzed related surveys in the context of smart grid and DT. 
  We notice the gap in lacking "survey" articles about DT-enhanced industrial cybersecurity. Despite it has been proposed in recent researches, none of them are in a "survey" form reviewing existing works and providing knowledge background needed by future developers.
  Therefore, our survey tries to fulfill the blank and introduce essential background knowledge covering SMART GRID, CYBERSECURITY, and DT to promote the development of DT-enhanced industrial cybersecurity.
  
  \item We introduce the background of smart grid to provide basic knowledge for interdisciplinary academics. 
  
  \item We review the last-decade cyberattack incidents in energy sectors and introduce nine prevalent attack methods.
  
  \item We introduce six critical defense approaches promising in protecting the smart grid from sophisticated cyber threats in both passive and active ways. These approaches include device identification, vulnerability discovery, intrusion detection, honeypot, attribution, and threat intelligence (TI). We indicate the possible collaboration between them and point out the challenges and future works for each technology.

  \item We review the existing works about DT, including its concept, components, applications in the smart grid, and DT as an enabler for enhanced cybersecurity. 

  \item We present our security considerations of the DT-based smart grid. The lessons learned and future perspectives are discussed from two aspects: i) Embedding DT into the security architecture of the smart grid, and ii) Deploying defense approaches for DT’s own security.

\end{itemize}

Section~\ref{Section: Related Surveys} reviews the related surveys in smart grid, cybersecurity, and DTs.
Section~\ref{Section: DT in the Smart Grid} introduces the background of the smart grid.
Section~\ref{Section: Cyber Attacks} reviews smart grid attack incidents and prevalent attack methods.
Section~\ref{Section: Defense} introduces critical defense approaches, including device identification, vulnerability discovery, intrusion detection, honeypot, attribution, and TI. 
Section~\ref{DT in SG} introduces DT, DT applications in the smart grid, and DT as an enabler for enhanced cybersecurity.
Section~\ref{Section: Digital Twin Enhanced Security-Lesson Learned and Future Perspective} discusses the lessons learned and future perspectives on the security considerations of DT-based smart grid.
Section~\ref{Section: Conclusion} presents the conclusion.

\section{Related Surveys}
\label{Section: Related Surveys}

The survey covers SMART GRID, CYBERSECURITY, DT, and their interactive part which represents DT-enhanced smart grid cybersecurity.
Thus, the paper firstly reviews and analyzes existing related surveys and tries to identify the research gap on this topic. To the best of our knowledge, the related surveys can be generally classified into two types. The first type focuses on the smart grid and its security issues. The second one targets the DT technology and its enabled applications.
TABLE~\ref{tab: Comparison of Related Surveys} has listed and analyzed the state-of-the-art surveys.
It can be observed that smart grid surveys mostly focus on the i) smart grid concept and components, ii) smart grid related technologies and applications, iii) smart grid communications and protocols, and iv) smart grid cybersecurity. Surveys of DTs mostly focus on the v) DT concept, development, and applications, but lack of a discussion about applications of the vi) DT in the smart grid, and vii) DT's security considerations. 
Therefore, our survey covers above topics to provide a systematic analysis of smart grid's security issues and fulfill the gap of lacking discussions about DT in smart grid's cybersecurity context.

\begin{table*}[htbp]
\centering
\caption{Summary of Related Surveys}
\label{tab: Comparison of Related Surveys}
\begin{tabular}{|l|c|ccc|c|cccc|c|c|}
\hline
\textbf{Year of   Publication}                        & \textbf{2016}                 & \multicolumn{3}{c|}{\textbf{2018}}                                                                  & \textbf{2019}                    & \multicolumn{4}{c|}{\textbf{2020}}                                                                                                  & \textbf{2021}                & \multirow{2}{*}{\textbf{Ours}} \\ \cline{1-11}
\textbf{Research   Area}                              & \textbf{\cite{tan2016survey}} & \textbf{\cite{faheem2018smart}} & \textbf{\cite{sun2018cyber}} & \textbf{\cite{tao2018digitaltwin}} & \textbf{\cite{musleh2019survey}} & \textbf{\cite{dileep2020survey}} & \textbf{\cite{hui20205g}} & \textbf{\cite{gunduz2020cyber}} & \textbf{\cite{minerva2020digital}} & \textbf{\cite{lo2021review}} &                                \\ \hline
Smart   Grid Concept and Components                   & $\bullet$                     & $\bullet$                       & $\bullet$                    & $\circ$                            & $\circ$                          & $\bullet$                        & $\circ$                   & $\bullet$                       & $\circ$                            & $\circ$                      & $\bullet$                      \\
Smart   Grid Related Technologies and Applications    & $\bullet$                     & $\bullet$                       & $\circ$                      & $\circ$                            & $\circ$                          & $\bullet$                        & $\bullet$                 & $\circ$                         & $\circ$                            & $\circ$                      & $\bullet$                      \\
Smart   Grid Communications and Protocols             & $\bullet$                     & $\bullet$                       & $\bullet$                    & $\circ$                            & $\circ$                          & $\bullet$                        & $\bullet$                 & $\circ$                         & $\circ$                            & $\circ$                      & $\bullet$                      \\
Smart   Grid Cybersecurity                            & $\bullet$                     & $\bullet$                       & $\bullet$                    & $\circ$                            & $\bullet$                        & $\circ$                          & $\bullet$                 & $\bullet$                       & $\circ$                            & $\circ$                      & $\bullet$                      \\
DT Concept, Development, and Applications & $\circ$                       & $\circ$                         & $\circ$                      & $\bullet$                          & $\circ$                          & $\circ$                          & $\circ$                   & $\circ$                         & $\bullet$                          & $\bullet$                    & $\bullet$                      \\
DT in the Smart Grid                      & $\circ$                       & $\circ$                         & $\circ$                      & $\circ$                            & $\circ$                          & $\circ$                          & $\circ$                   & $\circ$                         & $\circ$                            & $\circ$                      & $\bullet$                      \\
DT's Security Considerations                        & $\circ$                       & $\circ$                         & $\circ$                      & $\circ$                            & $\circ$                          & $\circ$                          & $\circ$                   & $\circ$                         & $\circ$                            & $\circ$                      & $\bullet$                      \\ \hline
\end{tabular}

Note: $\bullet$ area is involved in the survey, $\circ$ area is not involved.

\end{table*}

\subsection{Smart Grid and Security}

Dileep et al. \cite{dileep2020survey} introduced the background knowledge of the smart grid, including its definition, characteristics, functions, evolution, reference architecture, and components. Then, the authors reviewed the smart grid enabling technologies, i.e., smart meters, plug-in hybrid electric vehicle (PHEV), smart sensors, automated meter reading, vehicle to grid (V2G), and sensor and actuator networks. 
Further, the authors concluded the smart grid metering components, including advanced metering infrastructure (AMI), intelligent electronic devices (IEDs), and phasor measurement units (PMUs). As well, the communications of smart grid involving cloud, wide area network (WAN), wide area measurement systems (WAMS), neighborhood area network (NAN), home access network (HAN), and local area network (LAN) are introduced. 
Moreover, the authors discussed smart grid applications, including feeder automation, smart substation, and home and building automation. 
Their paper does not involve the security topics but is a great start for researchers and engineers to learn the smart grid and helps operators and authorities to build the smart grid and develop unified standards applicable for various applications. 

Faheem et al. \cite{faheem2018smart} introduced the smart grid in the context of "Industry 4.0". They reviewed the smart grid applications, including AMI, demand response (DR), substation automation, PHEV, Distributed energy resource (DER), etc. 
The authors also discussed the critical components in the smart grid, including IoT, CPS, big data, cloud computing, Internet of Services (IoS), cybersecurity, and communication technologies. It is worth noticing that they performed a general analysis of the cybersecurity issues. Security is also emphasized in their proposed future works demanding further efforts in improving the reliability, efficiency, and security of communication processes. However, cybersecurity is only a small section of their paper. They did not discuss it in detail.

Hui et al. \cite{hui20205g} aim at providing guidelines for developing 5G in the smart grid DR. They investigated related works of the DR in recent practical advances, cybersecurity, consumer privacy, and its reliability. Further, the authors discussed the 5G technology and presented the potentials and feasibility of applying 5G in the smart grid DR. The authors indicated that the massive connection, fast data transfer speed, low power consumption, high reliability, robust security, and privacy make 5G applicable for DRs.

Sun et al. \cite{sun2018cyber} focused on the cybersecurity of the power grid. The authors pointed out the vulnerabilities of firewalls in defining security-perfect detection rules and preventing attacks bypassing its protection. Besides, the lack of strong cryptographic protection for power grid communication protocols and devices also results in the power grid's vulnerabilities. Potential threats exist in the synchronization of smart grid data, the vulnerability of wireless communication, the validation of anomaly detection and intrusion detection systems (IDSs), coordinated attacks, and human factors. Although the authors reviewed both vulnerabilities and protection mechanisms (i.e., anomaly detection and IDSs in their survey), it is not in details nor comprehensive. 

Musleh et al. \cite{musleh2019survey} focused on a specific security issue in the smart grid, i.e., the false data injection (FDI) attacks. The authors divided FDI attacks into physical-based, communication-based, network-based, and cyber-based FDI attacks. They proposed the impacts of FDI attacks on the economy and stability. Moreover, they summarized the detection algorithms, including model-based detection and data-driven detection.

Tan et al. \cite{tan2016survey} reviewed the smart grid security vulnerabilities and solutions from the perspective of the data lifecycle. It is composed of "data generation, data acquisition, data storage, and data processing". Besides, the authors discussed security analytics applying big data to cybersecurity. The authors indicated that data-driven security analytics would enable intelligent services such as predictive capabilities and automated real-time controls.

Gunduz et al. \cite{gunduz2020cyber} analyzed threats and potential solutions of the smart grid. They proposed cybersecurity objectives as confidentiality, integrity, and availability (CIA triad). As well, they proposed cybersecurity requirements involving authentication, authenticity, authorization, accountability, privacy, dependability, survivability, and safety criticality. Moreover, the most significant contribution of their survey is presenting a comprehensive review of attacks and solutions with classifications according to the CIA triad and network layers. 

In general, existing surveys either focus on the smart grid itself or its security issues. 
  For the smart grid, its definition, architecture, components, communication technologies,   protocols, and applications are widely introduced. It helps to study the smart grid and guide its development. 
  For security issues, the security demands, smart grid vulnerabilities, cyberattacks, and defending technologies are reviewed. It provides the basic guidelines for the design of security mechanisms. 
However, most surveys lack of a summary of historical cyberattack incidents, attack meathods, and the defense approaches against sophisticated attacks such as advanced persistent threat (APT) for the purpose of breaking interdisciplinary knowledge barriers.
  The smart grid security protection should not be limited to detecting and preventing malicious activities but tracing the identity of adversaries to solve security issues from the source. 
Therefore, our survey analyzes smart grid cybersecurity issues with a comprehensive review of smart grid external attacks. 
Besides, the defense approaches introduced in our paper aim at providing a deep protection of smart grid systems to solve the most severe and prevalent problems, such as Distributed Denial of Service (DDoS) attacks and APT, and prevent them from the source.

\subsection{Digital Twin}

Tao et al. \cite{tao2018digitaltwin} summarized the applications of DT in the industry context. The authors first reviewed the proposals of DTs from 2003 to 2018 to explain the concept of DTs. Further, the authors introduced existing works in DT modeling and simulation, interaction and collaboration, data fusion, and services. 
Then, the authors introduced the application of DTs in prognostics and health management (PHM), production, product design, etc \cite{tao2018digitaltwin}. 
The authors emphasized the core position of modeling in DTs and presented that the PHM is the most popular application of DTs in the industry. Additionally, the authors presented two promising applications of DTs, which are dispatching optimization and operational control. 

Minerva et al. \cite{minerva2020digital} indicated that DTs had derived various approaches and requirements in different scenarios. Consequently, a general concept of DTs is needed to make it widely compatible. The authors reviewed the state-of-the-art that had defined and applied DTs in manufacturing, virtual reality (VR), multiagent systems, augmented reality (AR), and virtualization. Besides, the authors pointed out DT's key properties and features in the IoT context. They proposed that DT's potential applications in the IoT context include virtual sensors, the digital patient, the digital city, and the cultural heritage.

Lo et al. \cite{lo2021review} reviewed researches of DTs in product design and development. The authors concluded that the DT could simplify the design processes. It is of help for the product concept generation and redesign. Besides, with big data, cloud computing, VR, and AR technologies, the DTs could analyze the large volume of data generated during the whole product life cycle in the real environment and make the product design visible for verification.

In general, most of the existing surveys reflect the development status of DTs.
On the one hand, they reviewed the existing concept definition, architecture design, and applications of DTs. On the other hand, they proposed the functions, features, and properties that DTs should meet in future development. Usually, they would present their concerns on the challenges in developing DTs and indicate the promising applications of DTs in different scenarios. 
The purpose of these surveys is to discuss the current development status of DTs and guide the future development and application in academic and industrial fields. 
However, they lack a comprehensive summary on DT's applications in specific industrial context (e.g., smart grid) or neglect the works of DT as an enabler for enhanced cybersecurity.

\section{Background of the Smart Grid}

\label{Section: DT in the Smart Grid}

\begin{figure*}[htbp]
\centerline{\includegraphics[scale=0.9]{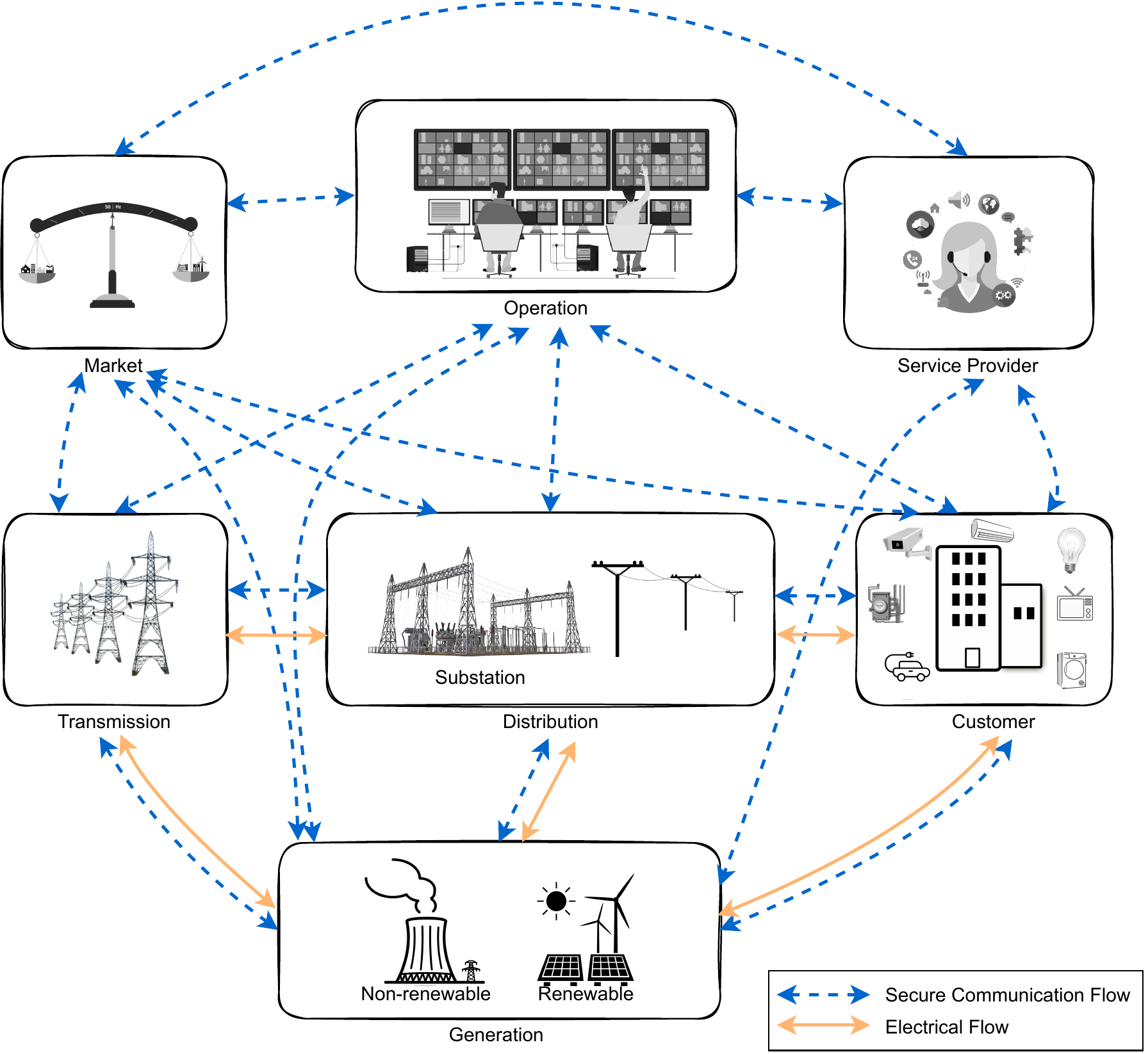}}
\caption{Conceptual model of the smart grid.}
\label{fig: Conceptual Model of Power Grid CPS}
\end{figure*}

The smart grid is a complicated system. The U.S. National Institute of Standards and Technology (NIST) presented a seven-domain infrastructure, including generation, distribution, transmission, markets, customers, service provider, and operations \cite{arnold2010nist,greer2014nist}. As illustrated in Fig. \ref{fig: Conceptual Model of Power Grid CPS}, 
the customer, generation, distribution, and transmission fields carry the energy flows. Meanwhile, all the seven elements are interconnected with mutual information channels for data interactions. 
Firstly, the generation domain converts different forms of energy (e.g., coal, nuclear, wind, solar and hydro) into electricity. It generally covers the traditional bulk generation and renewable power generation. The novel DER is also part of it but realized in a more complicated way. It involves solar photovoltaics, gas-fired distributed generation, small and medium-size wind farms, energy storage (ES), electric vehicles, and demand-side management\cite{perez2016transmission}. 
Afterward, the generated power flows into the civil customers and industry consumers through the transmission lines and substations deployed in the transmission and distribution domains under the control of the energy management systems (EMSs) and the distributed management systems (DMSs)\cite{he2016cyber}. Alternatively, the energy could also be saved in the ES for balancing the inflexible or intermittent supply with demand \cite{schmidt2017future}. 
Additionally, the management and control are mainly conducted in the operating systems (OSs) like the supervisory control and data acquisition (SCADA) system. It optimizes the grid energy according to the system status and power consumption reported by the terminal equipment (e.g., PMU) and remote terminal units (RTUs)) and the smart meters in AMI to achieve a balance between the supply and demand. Meanwhile, the OSs monitor the power grid to prevent anomaly behaviors. Once power grid failures or attacks are detected, it will assist security experts in emergency responses.
Similarly, the power grid status and the electricity consumption data will also be delivered to the energy market and service providers to adjust real-time prices and provide various intelligent services.


Besides, the smart grid has become one of the most representative CPSs. It realizes the deep integration of the cyber-physical world through communication, computer, and control (3C) technologies. More precisely, it involves the traditional information technology (IT) for data transmitting and the operational technology (OT) for control and actuation \cite{CPS2017NIST}. 
\begin{figure}[ht]
\centerline{\includegraphics[scale=1]{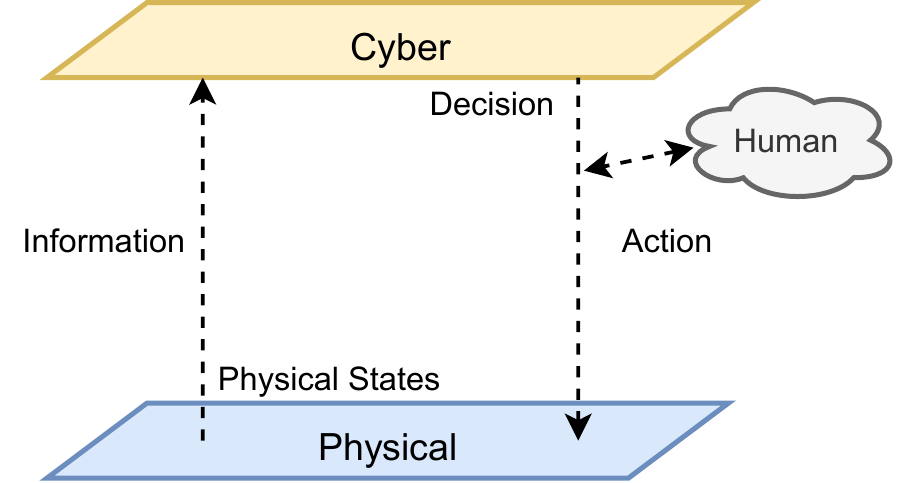}}
\caption{CPS conceptual model.}
\label{fig: CPS model}
\end{figure}
The interaction between the cyber and physical world in CPS has been illustrated in Fig. \ref{fig: CPS model}. The information of physical states is transmitted to the cyber side for decision-making, and the cyber system will send control commands to the actuation units to manipulate its physical states.
The whole process requires the coordination and cooperation of computing devices, communication networks, sensing or data acquisition devices, and the actuation units. As well, it provides the smart grid abilities of data fusion, distributed collaboration, real-time situational awareness, system adjustment, global optimization, and rapid emergency response.

Moreover, the development of modern information and communication technologies (ICTs) like DTs, big data, cloud, mobile communication, AI, and the IoT, will accelerate the digital and intelligent transformation of the smart grid. The deep integration of diversified ICTs will significantly reinforce the energy system's regulatory capabilities and promote the coordinated interaction among power generation, power grid, energy load, and ES. In addition, it will support the development of novel distributed energy techniques, micro-grids, electric vehicles, and any other energy applications to realize the modernization reform of the energy industry. 

However, each single step towards a digital smart grid also exacerbates its cybersecurity problems.
For smart grid terminal devices, the exposed interfaces, hardware design flaws, unpatched software bugs, backdoor, and default login password will all make them vulnerable to malicious adversaries. More seriously, the large volume of devices could aggravate the risks and breaks out with a whole different magnitude\cite{ylianttila20206g}. Once malicious adversaries successfully manipulate a large volume of devices such as programmable logic controllers (PLCs), smart meters, or RTUs, severe impacts can be added on the performance of the physical grid, electricity market, and customer services. 
A typical example is building botnets with compromised devices to perform DDoS attacks (introduced in Section~\ref{Section: Cyber Attacks}). 
It could result in the interruption of customer services, power outage, line overloads, and damage of critical smart grid infrastructures.
Besides, the smart grid has integrated various communication technologies, standards, and protocols. Wired communication like Ethernet, fiber optics, digital subscriber line (DSL), power line communication and wireless communication like 5G, Wi-Fi, DASH 7, Bluetooth \cite{faheem2018smart} have been applied for different smart grid scenarios. Dedicated standards and protocols like DNP3, IEC 61850, and IEC60870-5 are developed for specific power system communications. 
Their security vulnerabilities on transmitting with plaintext, lacking strong encryption and tamper protection, improper authorization, access control, and key management techniques have significantly increased the risks of the smart grid \cite{kumar2019smart}.
Yoo et al. \cite{yoo2016challenges} provided a study case of a smart grid environment in Korea with two substations and two hierarchical control centers (EMS/SCADA) using DNP3, IEC 61970, IEC 61850, and OPC UA. The authors clarified that the security threats exhibited in their study case include protocol vulnerability, improper security service mapping, improper protocol mapping, insecure gateway system, insecure configuration tool, and network design weakness \cite{yoo2016challenges}. 
All the vulnerabilities could result in severe damages to the smart grid. Cybersecurity measures need to be enhanced either by improving existing defensing methods or developing novel technologies, like DT, to ensure smart grid's resilience against cyberattacks.

\section{Smart Grid Cyber Attacks}

\label{Section: Cyber Attacks}

Adversaries will do their best to discover power system's vulnerabilities for illegally obtaining private information and economic benefits or, more severely, damaging the power grid facilities. In 2013, the European Union Agency for Network and Information Security (ENISA) published smart grid threat landscape to detailly describe the potential cyberattacks in the smart grid \cite{Threat2013enisa}. In 2020, ENISA presented the top 15 threats that occurred from January 2019 to April 2020, including “malware, web-based attacks, phishing, web application attacks, spam, DDoS, identify theft, data breach, insider threat, botnets, physical manipulation, damage, theft and loss, information leakage, ransomware, cyberespionage, and crytojacking \cite{Top52020Threats}”. Although the ENISA’s report did not specifically target the energy sectors, it fits the smart grid scenarios due to the application of ICT and the shift of adversaries’ focus from the general cyber environment to critical national infrastructures, e.g., the energy sectors. As a result, for better understanding the cyber threats of the smart grid, this section introduces the cyberattack incidents in power industries in the last decade. Additionally, this section summarizes the prevalent attack methods hurting energy sectors so that security experts can better protect the smart grid.

\subsection{Attack Incidents}
\begin{figure*}[htbp]
\centerline{\includegraphics[scale=0.7]{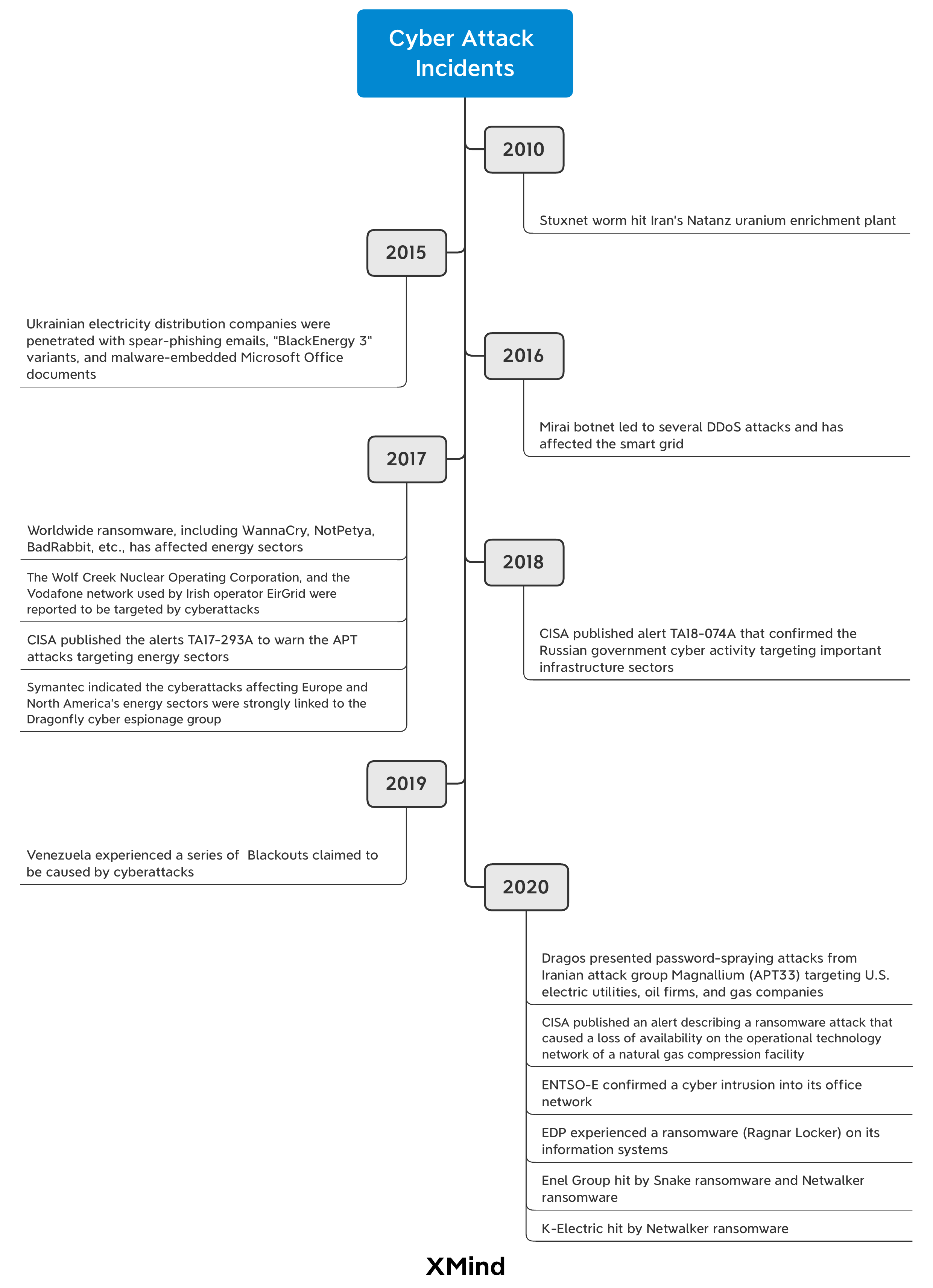}}
\caption{Timeline of Attack Incidents in the Smart Grid.}
\label{Timeline of Attack Incidents}
\end{figure*}
In 2010, Stuxnet, a malicious computer worm, was first uncovered, which caused severe damage to Iran’s Natanz uranium enrichment plant. It targeted the ICSs by infecting any Windows PC it could find and dropping rogue code to specific PLCs to sabotage the power facilities \cite{langner2011stuxnet,falliere2011w32}.

On December 23, 2015, cyberattacks targeted Ukrainian electricity distribution companies. Adversaries penetrated the electricity companies’ networks with spear-phishing emails, “BlackEnergy 3” variants, and malware-embedded Microsoft Office documents, causing several outages affecting around 225,000 customers \cite{case2016analysis,ICS2016Alert}.

In August 2016, the Mirai botnet was first identified by the whitehat security research group MalwareMustDie and was reported to be responsible for several DDoS attack incidents, including the September attack incidents that targeted the Brian Krebs website, cloud service provider OVH, French web host, and the October attack against service provider Dyn which affected hundreds of websites such as GitHub, Netflix, Twitter, and Reddit\cite{kolias2017ddos,antonakakis2017understanding}. The new trend of DDoS attacks utilizing the vulnerabilities and large volume of IoT devices has started and severely affected cyberspace's security.

In 2017, ransomware, including WannaCry, NotPetya, BadRabbit, etc., appeared global wide and had affected various industries, including finance, energy, healthcare, and universities. 

On July 6, 2017, The New York Times published that "hackers had been penetrating the computer networks of nuclear power stations and other energy facilities in the United States and other countries \cite{Hacker2017FBI}". "According to security consultants and the urgent joint report issued by the Department of Homeland Security and the Federal Bureau of Investigation, the Wolf Creek Nuclear Operating Corporation, which runs a nuclear power plant near Burlington, Kan., was targeted \cite{Hacker2017FBI}". The attack was considered to be related to APT actors and appeared to be a preparation for future attacks.

On August 07, 2017, a report from Irish Independent \cite{State2017Eirgrid} claimed that hackers gained access to a Vodafone network used by Irish operator EirGrid in the U.K. and compromised the routers used by EirGrid in Wales and Northern Ireland. Hackers can access the unencrypted communications of the companies by installing a wiretap on the system. In addition, there is the probability that the power grid systems are injected with malicious software, and the commercial customers’ information is leaked by transferring over the compromised network. According to the news, Vodafone and the National Cyber Security Centre regarded the activity as a “state-sponsored attack”.

On October 20, 2017, the Cybersecurity \& Infrastructure Security Agency (CISA) published technical alerts TA17-293A \cite{CISA2017APT} to warn the public about APT attacks against energy and other critical infrastructure sectors.
On the same day, Symantec indicated the cyberattacks affecting Europe and North America's energy sectors were strongly linked to the Dragonfly cyber espionage group \cite{team2017dragonfly}. The group's early campaign was between 2011 and 2014, originally targeting the defense and aviation companies in the Canada and U.S., and shifted to energy facilities in 2013 \cite{Dragonfly2014}. From 2015 to 2017, a new wave of cyber-attacks aiming to learn how energy facilities operate to access the operational systems for potential sabotage was detected and related to the Dragonfly group, named “Dragonfly 2.0” by Symantec. The reports from both CISA and Symantec presented the "tactics, techniques, and procedures (TTPs) used by APT actors, including open-source reconnaissance, spear-phishing emails (from compromised legitimate accounts), watering-hole domains, credential gathering, host-based exploitation, and the target on ICS infrastructure". The campaign was analyzed using the Lockheed-Martin Cyber Kill Chain model, including "reconnaissance, weaponization, delivery, exploitation, installation, command and control, and actions on the objective"\cite{killchain}.

On March 15, 2018, CISA renewed the alert and published TA18-074A that confirmed the Russian government cyber activity targeting important infrastructure sectors, such as aviation, energy, water, nuclear, commercial facilities, and other sensitive manufacturing sectors \cite{cert2018russian}.

In March 2019, Venezuela experienced a series of blackouts, including a nationwide blackout that lasted for a week \cite{BBC019Venezuela}. The government claimed that the blackouts were related to U.S.-involved cyberattacks, but no confirmed evidence has been given yet.

In January 2020, industrial cybersecurity company Dragos presented password-spraying attacks from the Iranian cyberattack group Magnallium, also known as APT 33, targeting U.S. electric utilities, oil firms, and gas companies \cite{Iranian2020wired,Dragos2020threat}.

On February 18, 2020, CISA published an alert describing a ransomware attack that caused a loss of availability on the OT network of a natural gas compression facility \cite{CISA2020ransomware}.

On March 9, 2020, the European Network of Transmission System Operators for Electricity (ENTSO-E) confirmed that its office network suffered a cyberattack\cite{Entsoe2020news}. According to ENTSO-E's report \cite{ENTSOE2020Intrusion}, risk assessment and contingency plans have been taken to prevent further attacks, but no detailed information about the cyberattack was given. 

On April 13, 2020, Energy company Energias de Portugal (EDP) experienced a ransomware attack (appeared to be Ragnar Locker) on its information systems. The attackers claimed that they obtained over 10TB of information from affected systems and demanded 1580 Bitcoin (around \$10 million) \cite{ZDnet2020EDP}.

In June 2020, Enel Group, an Italian multinational energy company, was hit by Snake ransomware \cite{Enel2020Snake}. Once again, in October 2020, Enel Group was infected with another ransomware named Netwalker. The Netwalker ransomware operators claimed to have obtained several terabytes of data from the company, asked bitcoins worth \$ 14 million, and threatened to leak the data if the money was not paid \cite{Enel2020Netwalker}. In September 2020, the Netwalker ransomware also attacked Pakistan's largest private power company K-Electric for blocking their billing and online services and asked \$3,850,000 worth of bitcoin \cite{Netwalker2020Kelectric}. 

On January 26, 2021, the U.S. CISA published an ICS advisory \cite{CISA2021Fujialert} that presented some high severity flaws of the SCADA/HMI products, including Tellus Lite V-Simulator (Versions before v4.0.10.0) and Server Lite (Versions before v4.0.10.0) made by Japanese electrical equipment company Fuji Electric \cite{CISA2021Fuji}. The flaws could provide attackers with chances to compromise the systems.

A timeline of cyberattack incidents mentioned in this survey has been illustrated in Fig.~\ref{Timeline of Attack Incidents}. In general, adversaries aim to obtain customer information through eavesdropping, get financial benefits through ransomware, penetrate and sabotage the smart grid through sophisticated attacks, including phishing emails, malware, etc. Attacks trying to destroy the smart grid are more complicated, requiring the cooperation of various types of attack approaches and turn to be state-sponsored. APT has become the most severe threat for smart grid entities. In this survey, we focus on the attacks that cause severe damage to the smart grid. Therefore, introductions about prevalent attack approaches are listed in the following part.

\subsection{Prevalent Attack Methods}
Cyber attacks are typically discussed from the perspective of CIA (confidentiality, integrity, and availability) requirements \cite{lee2010guidelines,mathas2020threat}. The explanation of CIA is listed as follows:

\begin{itemize}
  \item \textit{Confidentiality:}
  Attacks violating confidentiality could be adversaries illegally accessing unauthorized resources by eavesdropping, security mechanism bypass, illegal escalation of privileges, identity fabricating, etc.

  \item \textit{Integrity:}
  Attacks violating integrity will damage the consistency of data. Adversaries could illegally tamper or destroy the original stored or transmitted information to cause direct damages or hide their illegal behaviors for future intrusions.

  \item \textit{Availability:}
  Attacks violating availability will reject the regular usage of resources by legitimate users. Adversaries illegally consume the computing or communication resources of the target system so that it is unable to respond to the normal request of legitimate users. In addition, adversaries could also intercept the normal request to make the target service appears to be unavailable. 

\end{itemize}

In this section, prevalent attacks in energy sectors will be introduced. TABLE \ref{tab: Cyber attacks information attributes and proposed schemes} summarizes the involved researches on the smart grid cyberattacks and their impact on the CIA requirements.

\begin{table*}[htbp]
\centering
\caption{Cyber attacks, information attributes, and proposed schemes.}
\label{tab: Cyber attacks information attributes and proposed schemes}
\begin{tabular}{m{2cm}<{\centering}m{2cm}<{\centering}m{9.5cm}<{\centering}cc}
\hline
\textbf{Security   Issues}       & \textbf{Information  attributes}           & \textbf{Schemes}                                                                                                                                                                   & \textbf{Years} & \textbf{References}                        \\ \hline
FDI                              & Integrity                                  & Allocating optimal power to   transmit data over wireless channels for false data detection and injection   by solving a Bayesian Stackelberg game                                 & 2020           & \cite{ghosh2020defending}                  \\
FDI                              & Integrity                                  & Comparing the daily power   consumption and the sum of residuals with specified thresholds                                                                                         & 2021           & \cite{DBLP:journals/tdsc/BhattacharjeeD21} \\
FDI                              & Integrity                                  & Using moving target defense that   proactively perturbs branch susceptances to change system parameters against   knowledgeable adversaries                                        & 2020           & \cite{DBLP:journals/tifs/ZhangDYCC20}      \\
FDI                              & Integrity                                  & Modeling the attack impact and   presented the optimal attack triggering disruptive remedial actions by FDIs   with minimized remaining time                                       & 2017           & \cite{DBLP:journals/tifs/TanNFYKIG17}      \\
FDI-based  cyber topology attack & Integrity                                  & Stealthily adding small changes   to the LMP to make customers pay more                                                                                & 2017           & \cite{liang2017generalized}                \\ \hline
GPS   Spoofing                   & Integrity                                  & Detecting GPS spoofing based on   its C/No in physical layer and detecting the FDI   caused by GPS spoofing basing on the SE in upper layer & 2014           & \cite{fan2014cross}                        \\
Time   synchronization attack    & Integrity                                  & A first difference ML model to detect time synchronization attacks                                                                                          & 2017           & \cite{wang2017detecting}                   \\
Time   synchronization attack    & Integrity                                  & Detecting time synchronization   attacks with the three-phase model in unbalanced power systems                                                                                    & 2021           & \cite{delcourt2021time}                    \\ \hline
Impersontion   Attack            & Confidentiality                            & A cross-layer impersonation   attack in 4G networks named IMP4GT                                                                                                                   & 2020           & \cite{rupprecht2020imp4gt}                 \\
Impersontion   Attack            & Confidentiality                            & A D-FES method  for detecting   impersonation attacks from Wi-Fi network data                                                            & 2017           & \cite{aminanto2017deep}                    \\ \hline
Jamming                          & Availability                               & A self-healing communication   scheme against jamming attacks                                                                                                                      & 2015           & \cite{liu2015enabling}                     \\ \hline
Malware                          & Confidentiality, Integrity,   Availability & Summarized the features of 16   widespread IoT malware in the last decade                                                                                                          & 2019           & \cite{vignau201910}                        \\
Malware                          & Confidentiality, Integrity,   Availability & A novel malware rootkit, named   Harvey, targeting the PLC firmware of the smart grid CPS                                                        & 2017           & \cite{garcia2017hey}                       \\ \hline
Botnet                           & Confidentiality, Integrity,   Availability & A review of Mirai, the variant   evolution and the compromised devices                                                                                                             & 2017           & \cite{antonakakis2017understanding}        \\
Botnet                           & Confidentiality, Integrity,   Availability & Manipulation of demand via IoT   (MadIoT) attacks utilizing the compromised high-wattage IoT devices   constructed botnet to manipulate the total power demand.                    & 2018           & \cite{soltan2018blackiot}                  \\ \hline
DoS                              & Availability                               & A DoS attack in the AMI network, named puppet attack                                                                                          & 2016           & \cite{yi2016puppet}                        \\
DoS                              & Availability                               & A DoS attack on grid-tied solar   inverter                                                                                                                                         & 2020           & \cite{DBLP:conf/uss/BaruaF20}              \\ \hline
APT                              & Confidentiality, Integrity,   Availability & A review of APT techniques,   solutions and challenges                                                                                                                             & 2019           & \cite{alshamrani2019survey}                \\ \hline
Attacks   against AI             & NA                                         & Adverarial inputs and poisoned   model attacks                                                                                                                                     & 2020           & \cite{DBLP:conf/ccs/PangSZJVLLW20}         \\
Attacks   against AI             & NA                                         & Membership inference attack                                                                                                                                                        & 2017           & \cite{shokri2017membership}                \\
Attacks   against AI             & NA                                         & Model inversion attack                                                                                                                                                             & 2015           & \cite{fredrikson2015model}                 \\ \hline
\end{tabular}
\end{table*}

\paragraph{False Data Injection}
FDI violates the integrity of information. Adversaries illegally inject errors or false data to disturb the normal behaviors of power grid CPS. Tan et al. \cite{DBLP:journals/tifs/TanNFYKIG17} presented that FDI attacks on automatic generation control (AGC) could cause frequency excursion, making generators disconnecting with customers, even resulting in blackouts. The authors modeled the attack impact and presented the optimal attack triggering disruptive remedial actions by FDIs with minimized remaining time. 
Ghosh et al. \cite{ghosh2020defending} targeted the FDI attack in the SE, where the supervision and control data is transmitted on a wireless powered sensor network (WPSN). The network comprises a central controller (CC) and multiple sensor nodes (SNs), including active SNs sending system measurement and idle SNs only transmitting data of critical events. Among them, adversaries compromise a subset of idle SNs to inject false data. The authors presented that allocating optimal power to transmit data over wireless channels is critical to both protectors and adversaries to detect FDI or inject false data, respectively. As a result, the authors formulated the communication between CC and adversaries as a Bayesian Stackelberg game to solve optimal strategies. 
Liang et al. \cite{liang2017generalized} proposed an FDI-based cyber topology attack which is proved to be able to stealthily add small changes to the locational marginal price (LMP) to make customers pay more. Additionally, it disturbs the transactions of the energy market. It is proved to be effective in mess up with Australian Electricity market trading mechanisms.
Zhang et al. \cite{DBLP:journals/tifs/ZhangDYCC20} analyzed the moving target defense (MTD), which is used to prevent FDI attacks by proactively perturbing branch susceptances to change system parameters against knowledgeable adversaries. It concluded the conditions that MTD works even when the FDI is generated with former breach susceptances. 
Bhattacharjee et al. \cite{DBLP:journals/tdsc/BhattacharjeeD21} proposed a two-tier data falsification detection scheme in AMI of decentralized micro-grids. The first tier analyzes the harmonic to arithmetic mean ratio of daily power consumption to confirm an attack. The second tier verifies the data falsification if the sum of residuals is beyond specified thresholds.


\paragraph{Time synchronization Attack}
Services and functions of the power grid rely on the availability of globally synchronized measurement devices \cite{moussa2016security}. For time synchronization realized with the global positioning system (GPS), the lack of encryption and authorization mechanisms will allow adversaries to construct forged GPS signals to disturb the time synchronization process. Fan et al. \cite{fan2014cross} proposed a cross-layer GPS spoofing detection scheme towards PMUs, which detects GPS spoofing basing on its carrier-to-noise ratio (C/No) in the physical layer and detect the FDI caused by GPS spoofing basing on the SE in the upper layer.
Wang et al. \cite{wang2017detecting} introduced the concept of “first-difference” from econometrics and statistics to represent the residual of time-series data. It is used for training a first difference machine learning (ML) model to detect time synchronization attacks.
Delcourt et al. \cite{delcourt2021time} investigated the advantages of SE using a three-phase model instead of the direct-sequence model in detecting time synchronization attacks. Their work demonstrated the feasibility of detecting time synchronization attacks with the three-phase model in unbalanced power systems.

\paragraph{Impersonation Attack}
Impersonation attack is a way that adversaries camouflage to be the legitimate parties in a system or a network protocol \cite{Adams2005}. Leveraging the LTE vulnerability of missing integrity protection on the user plane, Rupprecht et al. \cite{rupprecht2020imp4gt} proposed a cross-layer impersonation attack in 4G networks named IMP4GT, which enables adversaries to impersonate the phone or network on the user plane to send and receive arbitrary IP packets despite any encryption.
Aminanto et al. \cite{aminanto2017deep} presented a deep-feature extraction and selection (D-FES) method composed of an unsupervised Auto Encoding (AE) feature extractor, a supervised feature selection, and a neural network classifier for detecting impersonation attacks from Wi-Fi network data.

\paragraph{Jamming}
Jamming attacks damage the power grid by affecting the data transmitting performance of the wireless channel. Delayed information of time-critical power grid units could result in improper situation awareness and wrong system operations to cause severe damages. 
Liu et al. \cite{liu2015enabling} proposed a self-healing communication scheme against jamming attacks in smart grid. It is realized with intelligent local controllers and a retransmission mechanism to ensure sufficient readings from smart meters when jamming happens.

\paragraph{Malware}
Malware represents the software with malicious purposes, including the virus, ransomware, spyware, worm, etc. Well-known malware targeting national-state ICSs includes the Stuxnet worm against Iranian nuclear facilities  \cite{langner2011stuxnet} and the BlackEnergy 3 malware against the Ukrainian electricity distribution companies\cite{case2016analysis}. They all caused severe damage to the electricity facilities and brought terrible influences to national security and people's lives.
Vignau et al. \cite{vignau201910} analyzed and summarized the features of 16 widespread IoT malware in the last decade, including Linux.Hydra, Psyb0t, Chuck Norris, Tsunami/Kaiten, Aidra, Carna, Linux.Darlloz, Linux.wifatch, Bashlite, Remaiten, Hajime, Mirai, Amnesia, BrickerBot, IoTReaper, and VPNFilter. 
Garcia et al. \cite{garcia2017hey} proposed a novel malware rootkit, named Harvey, targeting the PLC firmware of the power grid CPS. It enables adversaries to cause physical damage and large-scale failures by replacing legitimate control commands with malicious ones. Besides, it covers its malicious behavior by injecting operator-expected measurements observed from a system simulation result with the original legitimate control command.

\paragraph{Botnet}

\begin{figure}[bp]
\centerline{\includegraphics[scale=0.85]{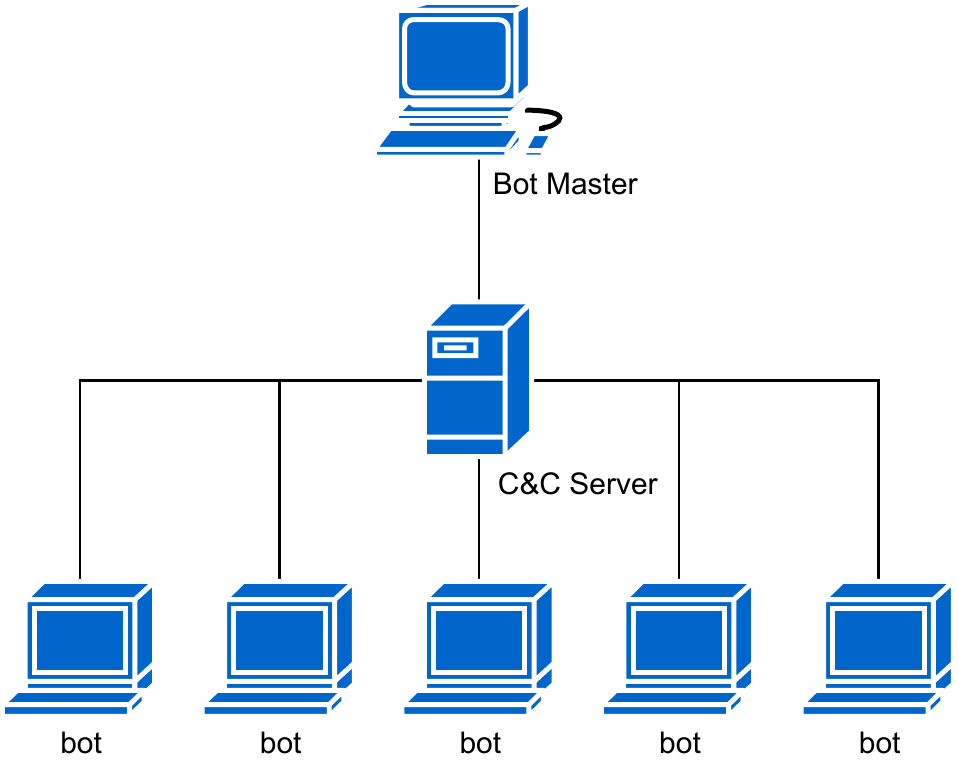}}
\caption{Typical centralized structure of botnet.}
\label{fig: botnet}
\end{figure}

Adversaries infect a number of host devices and communicate with them to obtain private information or perform malicious behaviors, e.g., DDoS. It embodies a network structure named botnet. Fig. \ref{fig: botnet} illustrated a typical centralized structure of botnet, which is composed of multiple compromised devices known as bots, a command and control (C\&C) server to manipulate bots, and the real adversary named bot master.
One of the most influential botnets, Mirai, has been introduced previously, which has a serious impact on cyberspaces’ security. 
Antonakakis et al. \cite{antonakakis2017understanding} reviewed the growth of Mirai to a peak of 600k infections in seven months and provided an analysis of Mirai's variant evolution, the compromised devices, and their concerns about the threat caused by large-scale IoT devices enabled botnet attacks.
Additionally, Soltan et al. \cite{soltan2018blackiot} demonstrated the feasibility of utilizing botnet to cause damages to the power grid as well as disturbing the electricity market. The authors presented that the Manipulation of demand via IoT (MadIoT) attacks targeted the load side of the energy supply. It utilizes the compromised high-wattage IoT devices (e.g., air conditioners, ovens, heaters, etc.) constructed botnet to manipulate the total power demand. It will result in the imbalance between energy supply and demand, bring in frequency instability, increase the operating cost, and, more severely, cause cascading failures and blackout.

\paragraph{DoS}
A denial of service (DoS) attack violates the availability of data. It exhausts the computing and communication resources of network nodes or the energy of target devices to make them inaccessible to their legitimate users. It results in the unavailability of measurements and makes power grid components out of service.
Yi et al. \cite{yi2016puppet} presented a DoS attack in the AMI network, named puppet attack. Adversaries select puppet nodes and send malicious packets to them. Correspondingly, puppet nodes will generate overmuch route packets to overload the communication bandwidth of AMI mesh networks, which results in a DoS.
Barua et al. \cite{DBLP:conf/uss/BaruaF20} demonstrated the feasibility of a DoS attack on grid-tied solar inverter. Adversaries could inject false measurements to spoof the Hall sensor of an inverter with an external magnetic field. The spoofing enabled adversaries to manipulate the inverter's output voltage, frequency, real and reactive power and result in grid failures. In addition, because the inverter is sensitive to the voltage variation, the overmuch voltage will turn off the inverter and result in a DoS.

\paragraph {APT}

\begin{figure*}[htbp]
\centerline{\includegraphics[scale=0.57]{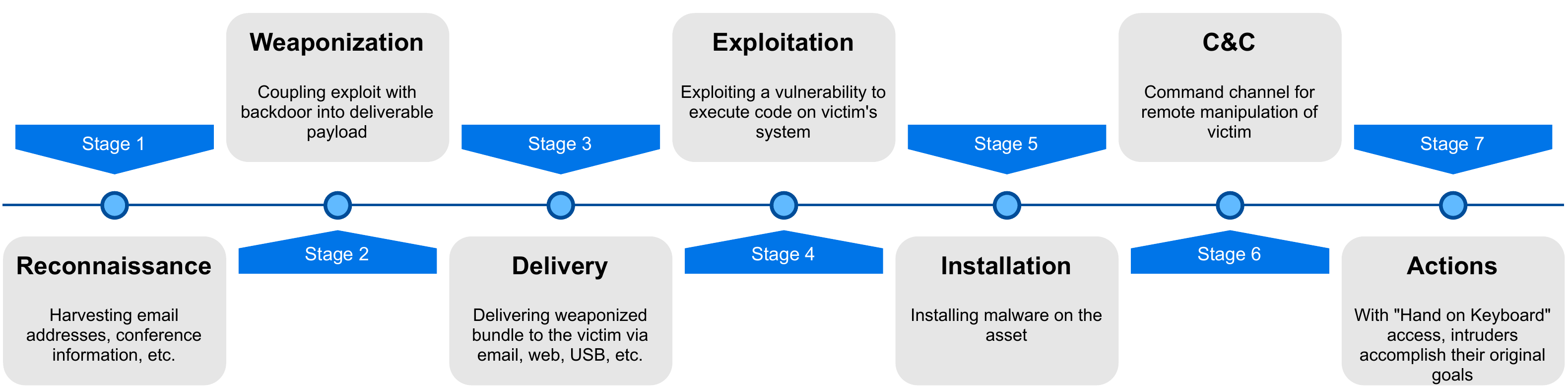}}
\caption{Lockheed-Martin cyber kill chain \cite{killchain}.}
\label{fig: kill chain}
\end{figure*}

Advanced Persistent Threat (APT) refers to a continuous attack activity carried out by a particular group or organization on a specific object. The APT groups could be nation-sponsored organizations with political and military purposes. The adversaries usually own rich resources and professional skills trying to perform concealed long-term penetration on specified targets. Their malicious behaviors could be generally explained by the seven-stage Lockheed-Martin kill chain model \cite{killchain} as illustrated in Fig. \ref{fig: kill chain}, including "reconnaissance, weaponization, delivery, exploitation, installation, command \& control, and actions on objectives". A more detailed description can be found in \cite{alshamrani2019survey}.

\paragraph {Attacks against AI}
\label{Section: Attacks against AI}

The realization of the DT-based smart grid relies on massive measurement and system status data. Its processing requires powerful analysis capabilities to meet the needs of intelligent services such as real-time energy regulation, safety supervision, and market analysis. AI has provided a practical solution due to its powerful data analysis capabilities. It is being applied to the overall scenarios of industry production and people's daily lives. As well, it plays a vital role in cyber defense. 
However, the widespread use of AI has also brought attention to its own security. Attackers began to study the vulnerability of AI algorithms and launch attacks against AI for security and privacy violations.
Pang et al. \cite{DBLP:conf/ccs/PangSZJVLLW20} discussed the adversarial inputs and the poisoned model attacks that significantly affect AI’s performance. For adversarial inputs, adversaries will modify the benign inputs into malicious ones causing ML models to make wrong predictions. For the poisoned model, adversaries will add malicious functions into the ML models. Then the models will trigger inappropriate behaviors when it receives adversary's pre-defined inputs. The authors analyzed the connections between these two attacks and then proposed a unified attack framework against AI models.
Shokri et al. \cite{shokri2017membership} proposed a membership inference attack to infer the private information contained in the training dataset by training shadow models to verify the difference of ML’s performance on training samples and first encountered samples.
Similarly, Fredrikson et al. \cite{fredrikson2015model} proposed a model inversion attack to infer sensitive features in the model inputs. Given a target label, the model inversion attack will start inferring its training input from an initial assumption and gradually add changes until the model prediction's confident value is high enough. Then the inferred input can be regarded as a similar copy of the original input, which is sufficient to expose the private information in it.


\section{Critical Defense Approaches}
\label{Section: Defense}

\begin{figure*}[htbp]
\centerline{\includegraphics[scale=0.8]{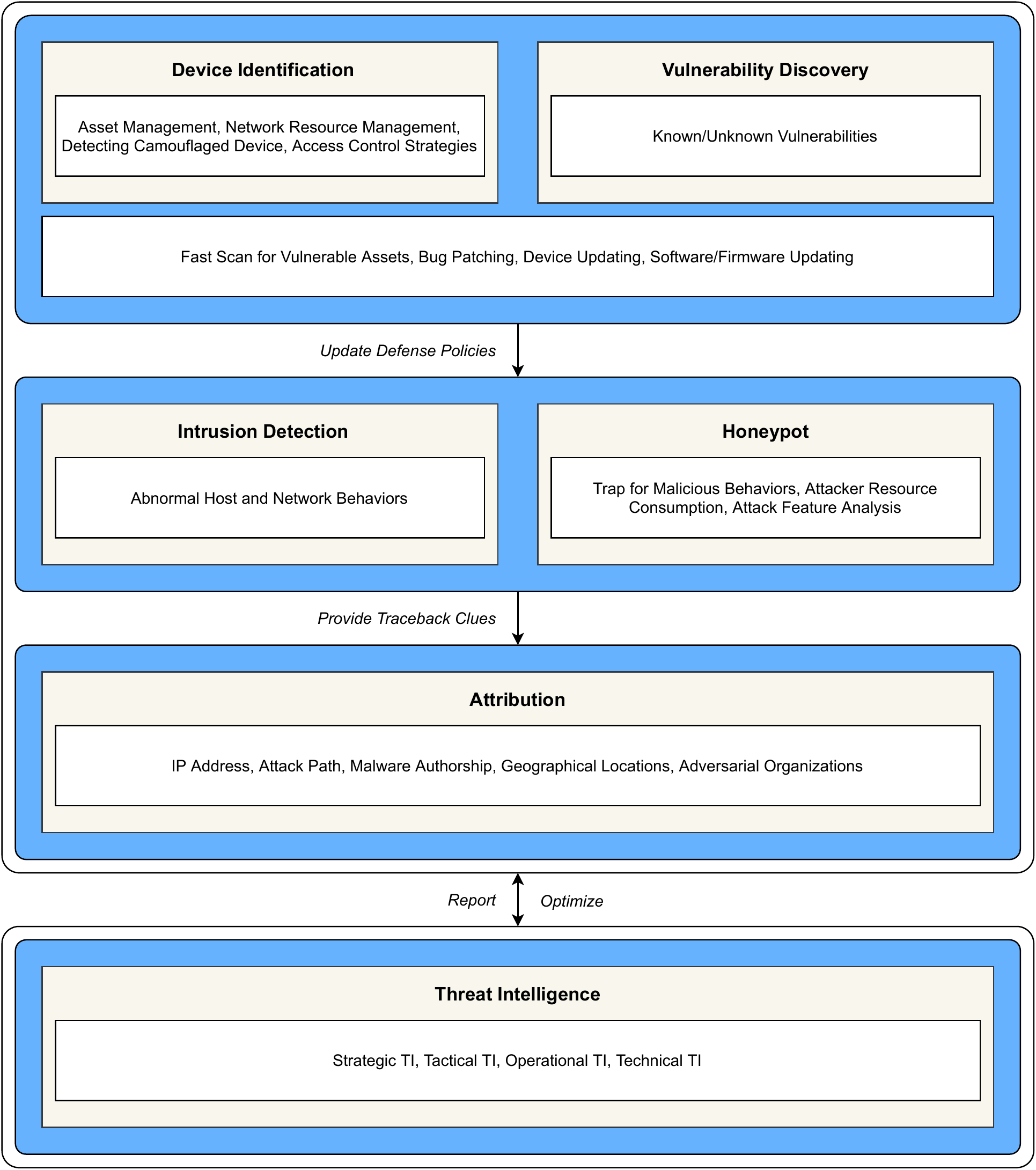}}
\caption{Cyber defense approaches}
\label{Figure: Cyber Defense Techniques}
\end{figure*}

To break the knowledge barriers between "SMART GRID, CYBERSECURITY, and DT", this section introduces the critical defense approaches that are promising in protecting industrial systems like the smart grid.
The involved technologies are illustrated in Fig.~\ref{Figure: Cyber Defense Techniques}. Firstly, we introduce device identification and vulnerability discovery approaches that target the vulnerabilities in physical devices and software to realize fast scan of vulnerable assets, bug patching, and software/firmware updating. Then, we review the IDSs targeting the vulnerabilities in devices, communications, and software to detect abnormal host and network behaviors. Moreover, honeypots, attribution, and TI are presented as an in-depth passive-active defense solution to prevent sophisticated attacks like APT.

\subsection{Device Identification}
\label{Section: Device Identification}



Large-scale heterogeneous IoT devices usually have different network resource and quality requirements in various smart grid scenarios. For example, smart kettles and smart meters vary in data collection and uploading capabilities. Devices should be allocated with different access authorities and network resources according to their device type. 
At this point, device identification is a possible solution. It recognizes the device types and helps to perform statistical and quantitative analysis of smart grid network assets. It benefits the formulation of network resource management and access control strategies specific to device types, realizing customized management of smart grid assets.

In addition, device identification is useful for the fast vulnerability discovery of large-scale smart grid devices.
Network device identification is conducive to the scanning of network assets. Security experts can maintain a vulnerability database where each bug is linked with its firmware and device type. When a new device is installed or connected into the smart grid network environment, network administrators could identify its device type, find out the firmware, and search for vulnerabilities if there exists any. Therefore, it can realize the quick vulnerability detection of large-scale smart grid devices and isolate suspicious equipment.

From the perspective of cyberattacks, device identification can figure out the granted identification of connected devices and the faithfulness of network access points. Thus, it is helpful to resist identity tampering and impersonation attacks. At the same time, device identification can formulate customized defense strategies based on the device type, such as firewall rules and IDS policies specific to device types.

In summary, device identification plays a significant role in smart grid network asset management. It is mainly reflected in the statistical analysis of network equipment assets, the configuration of network resources, the scanning and isolation of vulnerable equipment, the upgrade and maintenance of equipment firmware, and the formulation of customized defense strategies.

\subsubsection{Fingerprinting Techniques}
Existing device identification or fingerprinting methods include passive sniffing and active probing. The passive sniffing eavesdrops on the network traffic or radio frequency (RF) signals to extract device signatures. Differently, the active probing sends request packets to the target devices and extracts device features from the responding traffics. 
In general, both of the ways generate device signatures either from RF signals in the physical layer, traffic packets in the network, transport, and application layers, or their timing characteristics. 
Besides, the identification (or classification) of the device types is mostly realized on the AI algorithms. Nevertheless, a few approaches still match the extracted device signature with a pre-generated fingerprint database to realize device identification.

\begin{table*}[htbp]
\centering
\caption{Device Identification Techniques}
\label{tab:Device Identification Techniques}
\begin{tabular}{m{1.2cm}<{\centering}m{1.2cm}<{\centering}m{1.2cm}<{\centering}m{1.3cm}<{\centering}m{1cm}<{\centering}m{4cm}<{\centering}m{2.5cm}<{\centering}cc}
\hline
\textbf{Title}  & \textbf{Scope}  & \textbf{Target}        & \textbf{Granularity}             & \textbf{State}  & \textbf{Feature}                                                                                                                  & \textbf{Classification}                                     & \textbf{Year} & \textbf{Reference}                               \\ \hline
S\&F             & CPS             & CPS device             & Device                           & Active          & System and Function Calls,   Memory and CPU Utilization, Application Execution Time                                               & Matching Signature Database                                 & 2021          & \cite{babun2021cps}                              \\ \hline
OWL             & Wi-Fi           & Mobile and IoT Devices & Manufacturer, Type, Model        & Passive         & Protocol Attributes in Broadcast   or Multicast Packets                                                                           & MvWDL                                                       & 2020          & \cite{yu2020you}                                 \\ \hline
Audi            & SOHO Network    & IoT Device             & Device Type                      & Passive         & Timing Characteristics of   Periodic Background Communication Traffic                                                             & KNN                                                         & 2019          & \cite{marchal2019audi}                           \\ \hline
Yang et   al.   & Internet        & IoT Device             & Device Type, Vendor, Product     & Active          & Protocol Features from Network   Layer, Transport Layer and Application Layer                                                     & LSTM                                                        & 2019          & \cite{yang2019towards}                           \\ \hline
Yu et   al.     & ZigBee          & ZigBee Device          & Device                           & Passive         & Selected ROI for RF                                                                                                               & MSCNN                                                       & 2019          & \cite{yu2019robust}                              \\ \hline
SysID           & (-)             & IoT Device             & Device Type                      & Active, Passive & Features Selected by Genetic   Algorithm (GA) from Network Layer, Transport Layer and Application Layer   Protocols’ Header Field & DecisionTable, J48 Decision   Trees, OneR,  PART            & 2019          & \cite{aksoy2019automated}                        \\ \hline
PrEDeC          & Wi-Fi with WPA2 & IoT Device             & Device Type                      & Passive         & Header Information, Frame Size,   Timestamp of Link-layer Traffic Frames                                                          & Random Forest, Decision Tree,   SVM                         & 2017          & \cite{maiti2017link}                             \\ \hline
IOT   SENTINEL  & SOHO Network    & IP-enabled IoT Device  & Device Model, Software Version   & Passive         & Packet Features during The Setup   Phase between The Newly Introduced Device and Gateway                                          & Random Forest,   Damerau-Levenshtein Edit Distance Tiebreak & 2017          & \cite{miettinen2017iot}                          \\ \hline
Formby et   al. & ICS             & ICS Device             & Device and Software Type, Vendor & Passive         & Cross-layer Response Time, SER   Timestamp                                                                                        & FF-ANN, Bayes Classifier, GMM                               & 2016          & \cite{formby2016s}   \cite{gu2018fingerprinting} \\ \hline
GTID            & Local Network   & Wireless Device        & Device Type                      & Active, Passive & Timing Characteristics Caused by   The Difference of Hardware Compostions and Clock Skew                                          & ANN                                                         & 2014          & \cite{radhakrishnan2014gtid}                     \\ \hline
\end{tabular}
\end{table*}

\paragraph{Passive Fingerprinting}
Leveraging the control and data acquisition functions of SCADA protocols, Formby et al. \cite{formby2016s} proposed two device fingerprinting methods targeting ICS devices. First, regarding the data acquisition functions, the authors generated device and software type fingerprints according to the cross-layer response time (CLRT) between application layer response and TCP layer acknowledgement. It reflects the speed and workload of different IED devices and is related to the device hardware and software configuration. The second method regards the control functions of ICS devices. Devices with different mechanical and physical properties differ with the operation time, e.g., the time of a latching relay responding to an operating command. Therefore, the authors utilized the sequence of event recorder (SER) timestamp to represent devices’ physical features and realized the identification of device vendors. Moreover, the classification is processed by multiple machine-learning algorithms, including a feed forward artificial neural network (FF-ANN), a multinomial naive Bayes classifier, and Gaussian mixture models (GMM), on a live power substation and controlled lab experiments. In other words, it is practical for power grid device fingerprinting.
Maiti et al. \cite{maiti2017link} proposed an IoT device identification method named Privacy Exposing Device Classifier (PrEDeC). It is regarded as a method for attackers to obtain private information by passively eavesdropping on link-layer traffic of Wi-Fi encrypted by the Wi-Fi Protected Access 2 (WPA2) technique. It uses Scapy to extract features from PCAP files, and the features are generated from the header information, the frame size, and the timestamp of each traffic frame. Moreover, the authors trained Random Forest, Support Vector Machine (SVM), and Decision Tree models to realize the device type classification.
Miettinen et al. \cite{miettinen2017iot} aim to limit the communication of vulnerable devices so as to mitigate the security risks when adversaries compromise them. Thus, they proposed IoT SENTINEL to automatically identify device types in a home or small office network environment composed of IoT devices, wired or wireless network interfaces, and a gateway router. It passively collects traffic packets during the setup phases between newly introduced devices and the gateway. It extracts 23 types of fingerprint features from IP options, IP addresses, link layer, network layer, transport layer, and application layer protocols, packet contents, and port classes. In the experiment, IoT SENTINEL obtained a dataset of 540 fingerprints for 27 types of devices. Then, the identification combines two steps. Firstly, it trained a binary classifier with the Random Forest algorithm to tell whether the device features match any class or not. If the device features match several device types, it uses Damerau-Levenshtein edit distance to eventually decide the device identity. Furthermore, it queries repositories like the Common Vulnerabilities and Exposures (CVE) \cite{cve} database to check the existence of bugs for vulnerability assessment. Once a vulnerable device was detected, the security gateway built on the Software-defined Networking (SDN) architecture will constraint the communication and filter the traffic of detected vulnerable devices for risk control.
Marchal et al. \cite{marchal2019audi} proposed a passive identification system named AUDI. It automatically identifies the device type of IoT devices connected to the Small Office and Home (SOHO) network gateway. It extracts the timing characteristics of periodic background network traffic and uses the unsupervised clustering algorithm, KNN, to vary different device types.
Yu et al. \cite{yu2020you} proposed OWL (Overhearing on Wi-Fi for device identification), a mobile and IoT device identification method for Wi-Fi connections, which only depended on the traffic features extracted from passively received broadcast or multicast packets. The proposed method generated device fingerprints according to passively observed protocol attributes. Then the authors presented a multi-view wide and deep learning (MvWDL) algorithm for device identification and identified devices’ manufacturers, types, and models. 
Yu et al. \cite{yu2019robust} proposed a RF fingerprinting method for ZigBee devices. This method takes the physical lay features into account for device identification. It passively receives RF signals with a universal software radio peripheral (USRP). Considering the sleeping mode switching of ZigBee devices, the authors divide the received preamble signals into a semi-steady portion and a steady-state portion. Then, they apply an adaptive region of interest (ROI) selection algorithm to decide whether they need to filter out the semi-steady features for robust fingerprinting enhancement according to the signal-to-noise ratio (SNR). For feature extraction and device classification, the authors proposed a multisampling convolutional neural network (MSCNN), which downsamples the baseband signals into multiple time scales to improve the fingerprinting robustness.

\paragraph{Active Fingerprinting}
Yang et al. \cite{yang2019towards} presented that equipment manufacturers implement the network system on their products in different ways. To realize the identification of device type, vendor, and product, they proposed an active approach for automatic fingerprinting of IoT devices on the Internet. First, it actively sends query packets to remote hosts. In response, it extracts fingerprinting features of 20 protocols in the response packets from the network layer, transport layer, and application layer. Then it uses the vector transforming tool, Glove \cite{pennington2014glove}, to represent words into vectors and uses the neural network algorithm LSTM to process them and realize identification.
Babun et al. \cite{babun2021cps} proposed STOP-AND-FRISK (S\&F) as a host-based CPS device identification method to identify unauthorized or spoofed devices in the CPS environment. First, a remote server monitoring CPS environments will send a secure request to the unknown CPS device. Then, the device host will extract responding system/function calls (OS and kernel level features), CPU and memory utilization, and application execution time (hardware level features) to generate device signature. Finally, the generated signature will match a ground-truth device signature database to decide its identity.

\paragraph{Both Passive and Active Fingerprinting}
Radhakrishnan et al. \cite{radhakrishnan2014gtid} proposed GTID as a timing characteristic-based wireless device and device type identification system in local networks. It actively or passively detects devices according to the different distribution of packet inter-arrival time observed from the network's wire side, e.g., a backbone switch. The fingerprints are generated due to various hardware compositions and clock skews. It is identified by an Artificial Neural Network (ANN) model to complete the authentication, access control, network management, and counterfeit device detection schemes. The authors evaluated their approach in an isolated testbed as well as a live campus network consisting of 37 devices like iPhones, iPads, Kindles,  Netbooks, Google-Phones, etc. However, the timing characteristics will be lost in the buffering of switches or the routers. Therefore, this method is unsuitable for the Internet but adapts to the local network device identification.
Aksoy et al. \cite{aksoy2019automated} proposed an IoT device fingerprinting method named System IDentifier (SysID). It uses a single TCP/IP packet to identify device types with high accuracy. SysID collects protocol features from the network layer, transport layer, and application layer protocols’ header field and applies the genetic algorithm (GA) to filter out noisy features. The classification is processed by ML algorithms, including DecisionTable, J48 Decision Trees, OneR, and PART. The authors presented that their approach is suitable for both active fingerprinting and passive fingerprinting. They tested their system on 23 IoT devices in the IoT Sentinel dataset \cite{miettinen2017iot} and achieved over 95\% accuracy.

\subsubsection{Challenges and Future Works}
Device fingerprinting is meaningful in solving smart grid security issues brought from large-scale heterogeneous network devices. It helps identify vulnerable equipment and associates with other security mechanisms to build device-specific protection schemes. However, besides the original limitations of existing device fingerprinting methods, there are new challenges for the smart grid. The challenges and future works are summarized as follows.

\begin{itemize}
  \item \textit{Efforts for Smart Grid Device Fingerprinting:}
  According to our investigation, there are plenty of published researches, but only a few works have focused explicitly on the smart grid scenarios. More efforts should be made in future studies to develop smart grid-specific fingerprinting methods considering the inherent characteristics of power system equipment (e.g., packet features from explicit ICS protocols) to improve the accuracy in the critical electricity industry.

  \item \textit{Large-scale Heterogeneous Device Identification:}
  The realization of the smart grid with exploitations of DTs counts on the large-scale terminal devices for data acquisition. The collected information will converge into DT data and support the functions of DT entities. Therefore, the identification of large-scale heterogeneous power grid devices will become a novel challenge. It will place unprecedented requirements on the efficiency of device fingerprinting. Consequently, it is critical to introduce time overhead as a metric for fingerprinting evaluation. Fast and large-scale fingerprinting methods will be valued more than ever.

  \item \textit{Fine-grained Classification:}
  As illustrated in TABLE \ref{tab:Device Identification Techniques}, most researches focused on the identification of device type, manufacturers, and product models. On the one hand, it indicates that far more fine-grained device identification is hard to realize, e.g., identifying the software installed in physical devices. On the other hand, fine-grained identification accompanied by a large volume of identity labels brings the classification too much burden. As a result, for future works, it is necessary to discuss the specific identification purposes and security mechanisms before defining the fingerprint granularity. Moreover, the large volume of fingerprint labels reduces the value of accurate supervised learning-based classification because it requires a well-designed database which costs expensive human efforts to maintain. Thus, unsupervised learning will present its potentials in future works.

  \item \textit{Balance between Passive Fingerprinting and Active Fingerprinting:}
  Fingerprinting methods involve active probing and passive sniffing. Active fingerprinting achieves high accuracy by probing network devices with additional packets. However, power grid administrators might not be willing to take the risk of port scanning, especially for devices executing critical tasks as well as the vulnerable legacy devices \cite{formby2016s}. On the other hand, passive fingerprinting usually needs additional equipment (e.g., USRP) for network sniffing. It is far more expensive. Therefore, researchers should balance the choices between passive and active fingerprinting approaches.

  \item \textit{Phases of Identification in Devices’ Life Cycle:}
  Device fingerprinting could be processed during the device access phase, resource interaction phase, or the whole process in the devices’ life cycle. Fingerprint features vary from different stages. Therefore, researchers should determine their target before selecting identification approaches.

  \item \textit{Feature Selection:}
  In general, widely utilized features are extracted from the RF signals, network traffic, and timing characteristics. It varies from the physical layer, network layer, and application layer. In order to realize a fine-grained identification of heterogeneous power grid devices, it is necessary to merge features from different aspects and select proper ones for accurate fingerprinting. Meanwhile, it is worth noting that power grid IoT devices usually have less traffic data comparing to ordinary terminal devices. It samples and sends electrical data with a specific period. Devices work in a sleeping mode to save energy. Therefore, identification methods that require dense data will no longer be applicable. Future works need to realize the identification with fewer traffic packets.

\end{itemize}

In summary, future works should focus on the efficiency and granularity of device fingerprinting methods and make more efforts on the identification of smart grid-specific devices.

\subsection{Vulnerability Discovery}
\label{Section: Vulnerability Discovery}

The software or firmware installed in the massive smart grid devices might hold harmful program vulnerabilities. Attackers could exploit them to penetrate the power systems and access DT models to steal sensitive information, tamper with critical data, and forge control commands to affect the safety and reliability of the power grid systems or DT entities. Therefore, the vulnerability discovery approaches are essential in identifying vulnerable devices, patching detected bugs, and managing smart grid assets.

\subsubsection{Vulnerability Discovery Technologies}
Vulnerability discovery technologies are sorted in different ways. Based on the analysis object, they can be classified into source code analysis and binary analysis. Depending on whether the program is running, detection methods can be considered as static and dynamic analyses. Additionally, from the perspective of analysis methods, software analysis includes symbolic execution, fuzzing, taint analysis, ML-based analysis, etc. 
This section reviews the recently published or widely used vulnerability discovery works. It is expected to provide possible solutions in detecting vulnerabilities in the smart grid context.

\begin{table*}[htbp]
\centering
\caption{Summary of Vulnerability Discovery Researches}
\label{tab:Summary of Vulnerability Discovery Researches}
\begin{tabular}{ccccccc}
\hline
\textbf{Name or   Authors} & \textbf{Technique}              & \textbf{Object} & \textbf{State}  & \textbf{Year} & \textbf{Domain}          & \textbf{Reference}                       \\ \hline
Neutaint                   & DTA, Neural Network             & Binary          & Dynamic         & 2020          & General                  & \cite{she2020neutaint}                   \\ \hline
Sebastian et al.           & Symbolic Execution (Survey)     & -               & -               & 2019          & General                  & \cite{poeplau2019systematic}             \\ \hline
Roberto et al.             & Symbolic Execution (Survey)     & -               & -               & 2018          & General                  & \cite{baldoni2018survey}                 \\ \hline
KLEE                       & Symbolic Execution              & Source Code     & Static          & 2008          & General                  & \cite{cadar2008klee}                     \\ \hline
S2E                        & Symbolic Execution              & Binary          & Static          & 2012          & General                  & \cite{chipounov2012s2e}                  \\ \hline
angr                       & Symbolic Execution              & Binary          & Static          & 2017          & General                  & \cite{Angr}\cite{shoshitaishvili2016sok} \\ \hline
QSYM                       & Symbolic Execution              & Binary          & Static          & 2018          & General                  & \cite{yun2018qsym}                       \\ \hline
AFL                        & Fuzzing                         & Binary          & Dynamic         & 2013          & General                  & \cite{AFL}                               \\ \hline
AFL++                      & Fuzzing                         & Binary          & Dynamic         & 2020          & General                  & \cite{fioraldi2020afl++}                 \\ \hline
AFLFast                    & Fuzzing                         & Binary          & Dynamic         & 2017          & General                  & \cite{bohme2017coverage}                 \\ \hline
Firm-AFL                   & Fuzzing                         & Binary          & Dynamic         & 2019          & IoT Firmware             & \cite{zheng2019firm}                     \\ \hline
Wang et al.                & Fuzzing, Reinforcement Learning & Binary          & Dynamic         & 2021          & General                  & \cite{reinforcement2021wang}             \\ \hline
MOPT                       & Fuzzing                         & Binary          & Dynamic         & 2019          & General                  & \cite{lyu2019mopt}                       \\ \hline
Vuzzer                     & Fuzzing                         & Binary          & Dynamic         & 2017          & General                  & \cite{rawat2017vuzzer}                   \\ \hline
GREYONE                    & Taint-guided Fuzzing            & Binary          & Dynamic         & 2020          & General                  & \cite{gan2020greyone}                    \\ \hline
Angora                     & Taint-guided Fuzzing            & Binary          & Dynamic         & 2018          & General                  & \cite{chen2018angora}                    \\ \hline
VulDeePecker               & Deep Learning                   & Source Code     & Static          & 2018          & General                  & \cite{li2018vuldeepecker}                \\ \hline
$\mu$VulDeePecker          & Deep Learning                   & Source Code     & Static          & 2019          & General                  & \cite{zou2019muvuldeepecker}             \\ \hline
SySeVR                     & Deep Learning                   & Source Code     & Static          & 2021          & General                  & \cite{li2021sysevr}                      \\ \hline
Devign                     & Graph Neural Network            & Source Code     & Static          & 2019          & General                  & \cite{DBLP:conf/nips/ZhouLSD019}         \\ \hline
Karonte                    & Static taint analysis           & Multi-binary    & Static          & 2020          & IoT Firmware             & \cite{redini2020karonte}                 \\ \hline
SaTC                       & Symbolic Execution              & Multi-binary    & Static          & 2021          & IoT Firmware             & \cite{chen2021sharing}                   \\ \hline
Ying et al.                & Pattern Matching                & Source Code     & Static          & 2019          & Power Grid               & \cite{ying2019detecting}                 \\ \hline
Yoo et al.                 & Fuzzing, DTA                    & Binary          & Dynamic         & 2016          & Power Grid               & \cite{yoo2016grammar}                    \\ \hline
BinArm                     & Matching                        & Binary          & Dynamic         & 2018          & Power Grid               & \cite{shirani2018binarm}                 \\ \hline
EVA                        & Symbolic Execution              & Binary          & Static, Dynamic & 2017          & Power Grid               & \cite{kwon2017automated}                 \\ \hline
\end{tabular}
\end{table*}

\paragraph{Dynamic Taint Analysis (DTA)}
She et al. \cite{she2020neutaint} stated that the existing DTA methods follow the taint propagation rules to spread taint labels from source to sink. However, it is hard to draft rigorous taint propagation rules describing the overall situations precisely. DTA may suffer a high false alarm rate and the running time overhead for dynamic analysis. Consequently, the authors choose to embed a neural network model into the program to describe the information flows between source-sink pairs. With the assistance of a saliency map, the proposed system Neutaint is able to find out the sensitive input bytes that have the most significant impact on the output values. Subsequently, it could be used to guide fuzzing tools for better performance.

\paragraph{Symbolic Execution}
Symbolic execution represents program inputs and variables in symbolic format. Each path holds a set of constraints. When the path ends or triggers a bug, the constraint solver will figure out the concrete input space to reach the branch. Poeplau et al. \cite{poeplau2019systematic} and Baldoni et al. \cite{baldoni2018survey} summarized the existing symbolic execution techniques, including well-known KLEE \cite{cadar2008klee}, S2E \cite{chipounov2012s2e}, angr \cite{Angr}, and QSYM \cite{yun2018qsym}, etc. Among them, KLEE faces source code, while S2E, angr, and QSYM deal with binaries.

\paragraph{Fuzzing}
Fuzzing is a crucial method for vulnerability detection. Generally, it generates various program inputs to explore as many program traces as possible. By monitoring the dynamic program performance, fuzzing tools could discover the potential vulnerabilities. 
Currently, plenty of outstanding researches has been published. One of the most influential works is American fuzzy loop (AFL), which realizes instrumentation in compile-time and applies the genetic algorithm to generate valuable test cases triggering new program behaviors \cite{AFL}. 
It inspired many other consecutive works like AFL++ \cite{fioraldi2020afl++}, AFLFast \cite{bohme2017coverage}, Firm-AFL \cite{zheng2019firm}, etc. In addition, many researchers dedicated to enhancing fuzzer's performance in different ways.
Most fuzzers use coverage as the metric to guide seed selection. However, sensitive or fine-grained coverage metrics could select overmuch seeds, exceeding the fuzzer's scheduling ability and resulting in the seed explosion problem. Therefore, Wang et al. \cite{reinforcement2021wang} proposed a hierarchical seed scheduling method based on reinforcement learning to overcome seed explosion.
Lyu et al. \cite{lyu2019mopt} proposed an optimized mutation scheduling scheme MOPT to enhance the efficiency of fuzzers for generating valuable test cases. For mutation-based fuzzers, the mutation operator defines the mutation rules about where to mutate (which byte) and how to mutate (e.g., adding, removing, or replacing bytes). Mutation scheduling is the mutation operator selecting choices during different phases of fuzzing. The selection follows a probability distribution, and the purpose of \cite{lyu2019mopt} is to find the optimal one to generate more valuable mutations. MOPT applied the particle swarm optimization (PSO) algorithm to figure out probability distribution's optimal solutions. Additionally, the authors applied MOPT to well-known fuzzers, including AFL \cite{AFL}, AFLFast \cite{bohme2017coverage}, and VUzzer \cite{rawat2017vuzzer}. As expected, MOPT-based fuzzing obtained better performance (e.g., MOPT-AFL discovered 170\% more security vulnerabilities and 350\% more crashes).
Besides, taint analysis and symbolic execution are often used to guide fuzzers. For example, Gan et al. \cite{gan2020greyone} presented a taint-guided fuzzing method named GREYONE. GREYONE is composed of four parts, including fuzzing-driven taint inference (FTI), taint-guided mutation, core fuzzing, and conformance-guided evolution. Firstly, the FTI module infers the taint by monitoring variables’ value changes caused by input byte mutation. If the value of variables changes with the mutation of input bytes, it implies that the variable is tainted. Then, the taint inferred from FTI will guide the input byte mutation for generating fuzzing test cases. It optimized fuzzing speed by prioritizing the mutation of input bytes that influence more untouchable paths and the exploration of paths affected by more input bytes. Lastly, the authors presented constraint conformance as a data flow feature to select suitable seeds for mutation and explore new paths. GREYONE combined taint analysis with fuzzing to guide the evolution direction of fuzzing and improved the mutation efficiency.
Chen et al. \cite{chen2018angora} presented that fuzzers using symbolic execution to solve path constraints will consume too much time. 
Therefore, Chen et al.\cite{chen2018angora} provided a mutation-based fuzzing method in solving path constraints by tracking taints in the byte-level, counting branches, and exploring input length.

\paragraph{Deep Learning-based Vulnerability Discovery}
From 2018 to now, a research team published several deep learning-based methods \cite{li2018vuldeepecker,zou2019muvuldeepecker,li2021sysevr} for the detection of Library/API function call related vulnerabilities in C or C++ source code.
Li et al. \cite{li2018vuldeepecker} focused on detecting vulnerabilities caused by improper usage of Library/API function calls, such as resource management errors and buffer errors. The authors indicated that software vulnerabilities should not be analyzed without considering the program context. AI algorithms capable of processing program context should be preferentially recommended for the task of vulnerability detection. Due to the similarity of source code analysis and natural language processing (NLP), ideas or algorithms in the NLP field could be borrowed and applied to software analysis. Besides, the authors pointed out that to precisely identify the vulnerability locations, the granularity of software analysis should not be limited to the program or function level. Therefore, the authors proposed “code gadget” as the basic classification unit for AI-based vulnerability detection. Essentially, the code gadget is the extracted code fragments that have semantic relationships in the initial source code. In order to form code gadgets, they firstly identified Library/API function calls among the original code. Then, they generated data dependency graphs with the help of the commercial tool Checkmarx. Based on the graphs, they reorganized the program statements that share the same data flow with the identified Library/API function calls and formed code gadgets. Finally, the code gadget was vectorized and used as the input of the BLSTM model to identify whether there exist vulnerabilities in the program.
Zou et al. \cite{zou2019muvuldeepecker} improved their previous approach and realized multi-classification for software vulnerabilities. Differently, inspired by the concept of “region attention” (image fields providing more information for classification) in the image processing field, the authors proposed “code attention” indicating program statements that provide more valuable information for vulnerability detection. Specifically, for vulnerabilities caused by improper usage of Library/API function calls, program statements like Library/API function call, parameter definition, and control conditions should reveal more clues about the vulnerability. Therefore, the system formed code slices based on the three syntax features above to generate code attentions. Besides, based on the work in \cite{li2018vuldeepecker}, the authors added control-flow features into the original code gadget generation scheme. The ML model added one more BLSM module to process the features from both code gadgets and code attention. Then, an additional merge module will combine them for better classification.
Similarly, Li et al. \cite{li2021sysevr} were inspired by the concept of “region proposal” in image processing and presented that the key point of detection is to extract interesting regions in the code (vulnerable code segment) for analysis. The system extracts syntax features (such as function call, array usage, pointer usage, arithmetic expression, etc.) based on Abstract Syntax Tree (AST). In addition, the system extracts program slices with semantic features according to the program dependency graphs generated by the open-source tool, Joern. Consequently, both the syntax features and the semantic features are considered in the software analysis system and used as the input for different AI classification models.

\paragraph{Graph-based Vulnerability Discovery}
Zhou et al. \cite{DBLP:conf/nips/ZhouLSD019} presented the vulnerability detection approaches that process programs as flat text sequences using NLP algorithms can only represent partial features of the original programs. However, the program statements themselves possess far more complicated structures and logic features. Consequently, Zhou et al. \cite{DBLP:conf/nips/ZhouLSD019} proposed Deep Vulnerability Identification via Graph Neural Networks (Devign) to identify the existence of vulnerabilities in source code functions. It includes a Graph Embedding Layer, Gated Graph Recurrent Layer, and the Conv Layer. The Graph Embedding layer represented the source code functions as a joint graph, which combined Data Flow Graph (DFG), Control Flow Graph (CFG), AST, and Natural Code Sequence (NCS) for exhaustive representation of program semantic features. The Gated Graph Recurrent layer learns the characteristics of nodes by gathering and transferring information about neighboring nodes in the graph. Then, the Conv layer extracts node representations for graph-level prediction to identify vulnerabilities.

\paragraph{IoT Firmware Analysis}
Redini et al. \cite{redini2020karonte} indicated that embedded devices are composed of interconnected components like binary executable files or modules of a giant embedded OS. Different components communicate and cooperate with each other to finish various tasks. Attackers’ input from outside the network will not only affect binary files directly facing the network, but also other binary files. Any analysis that only focuses on these network-oriented binary files will omit the vulnerabilities in other binary files, resulting in a high false-negative rate (FNR). Therefore, the authors proposed a multi-binary vulnerability detection method, Karonte, targeting embedded device firmware. Karonte uses static analysis to link data-connected functions through multi-binary files so that tracking data flow through binaries is possible. Vulnerabilities crossing binary files triggered by the attacker's input can then be discovered by Karonte.
Chen et al. \cite{chen2021sharing} leveraged the common input keywords shared by frontend and backend binaries in embedded devices to fasten the locating of backend program statements processing user input data. Hereafter, vulnerabilities caused by user input can be analyzed more effectively by the proposed static taint checking system, SaTC, which is developed on the symbolic execution tool angr \cite{Angr}. The authors also compared SaTC with Karonte and claimed that SaTC discovered more bugs in embedded systems.

\paragraph{Smart Grid-specific Vulnerability Discovery}
Currently, vulnerability detection methods specific to power grid scenarios still account for a small proportion.
Ying et al. \cite{ying2019detecting} proposed a static source code analysis method for smart grid devices to detect buffer overflow vulnerabilities by matching extracted features with pre-defined vulnerability patterns.
Yoo et al. \cite{yoo2016grammar} proposed a grammar-based fuzzing method for SCADA systems. It analyzes programs with DTA to find out input satisfying the dependency relationships within execution paths and grammar constraints for SCADA’s protocols like Modbus.
Shirani et al. \cite{shirani2018binarm} proposed BINARM aims to detect vulnerabilities of IED firmware with ARM architecture. It firstly maintains a database with IED firmware and vulnerabilities by identifying various IED manufacturers, collecting their provided IED firmware, identifying used libraries, and looking for related CVE vulnerabilities. Then, the system matches the target firmware with the database to find out vulnerable functions.
Targeting AMI and EV charging systems in ARM architecture, Kwon et al. \cite{kwon2017automated} proposed a binary analysis method combining static analysis and dynamic analysis to discover security-critical vulnerabilities.

\subsubsection{Challenges and Future Works}
The vulnerability detection of the power grid involves software and firmware analysis of traditional computer systems, ICS, and power grid terminal IoT devices. The challenges and future works can be summarized as follows.

\begin{itemize}
  \item \textit{Lack of Dedicated Tools: }
  Genetic vulnerability detection tools are not dedicated to analyzing power systems. Developers are not familiar with power grid device functions, program features (e.g., inline assembly), code libraries, and potential bugs. Vulnerability detection methods designed for general purposes may have an excessively high FNR while analyzing software of power grid equipment. Therefore, vulnerability detection techniques targeting power systems should be developed and consider the specific features of vulnerable power grid software during the development and training processes.

  \item \textit{Source Code Hard to Obtain: }
  The source code of IoT device firmware is usually hard to obtain. The firmware published by device manufacturers is usually a set of binaries. Analyzers collect device firmware through web crawling or downloading from the manufacturer's official websites, but in most cases, further works can only be done on the basis of binary files or intermediate representation (IR). As a result, despite source code can provide more detailed syntax and semantic features, it will be binary analyzing techniques playing an essential role in power grid IoT device analysis.

  \item \textit{Binary Analysis: }
  Most binary analyzing works focus on detecting vulnerabilities in a single binary file. However, IoT firmware is usually constituted by several binaries. Different binary files share data to perform the tasks of the device. Nilo et al. \cite{redini2020karonte} presented that vulnerabilities triggered by the malicious input from external sources, e.g., through the network, may affect other binary files that are not directly facing the network. Therefore, the analysis only focusing on a single binary file will produce an unacceptable number of false alarms. However, in the 29 works listed in TABLE \ref{tab:Summary of Vulnerability Discovery Researches}, only two of them studied the multi-binary or cross-binary vulnerability detection methods, which is far from enough. As a result, future researchers should make more efforts to explore the vulnerabilities across multi-binaries. Moreover, for vulnerability detection methods that match firmware with vulnerability database, it is also essential to collect comprehensive firmware and power system vulnerabilities to form a firmware-vulnerability database.



  \item \textit{Evaluation of Detection Tools: }
  Due to the lack of a benchmark dataset of software vulnerabilities, existing works evaluate their detection tools by how many CVE bugs they could find. However, different experiment setups could lead to different results. There is no fair way to compare different detection tools. Besides, this method cannot tell how many bugs will the detection method miss. In other words, it cannot reflect their FNR. To the best of our knowledge, vulnerability generation could be a hopeful solution by artificially generating a vulnerability dataset. Future works could follow the idea to construct a standardized test dataset of power systems and evaluate the performance of various detection tools.

\end{itemize}

\subsection{Intrusion Detection}

Intrusion detection systems (IDSs) are applied to monitor network or computer system events, discover signs of security issues, and generate alarms when suspicious activities are detected \cite{liu2018host,bace2001nist}. 
Malicious activities failed to be prevented by the firewalls, access control, and authentication mechanism in the first place need to be detected by the IDS. Once cyber threats are discovered, they will be mitigated with pre-established security plans. 
Besides, IDS could cooperate with other security mechanisms.
With device identification, it is feasible to make device type-specific detecting rules to improve IDS’s accuracy. 
With honeypots, the IDS could assign the original malicious activities to a virtual environment to deceive attackers keeping interactions so that analysts could better extract attack characteristics and analyze their purposes. 
Additionally, IDS is an important source in generating TI. Conversely, TI could also update IDS rules to enhance its performance. 

This section introduces the IDS technologies and discuss the challenges and future works in the smart grid.

\begin{figure*}[ht]
\centerline{\includegraphics[scale=0.7]{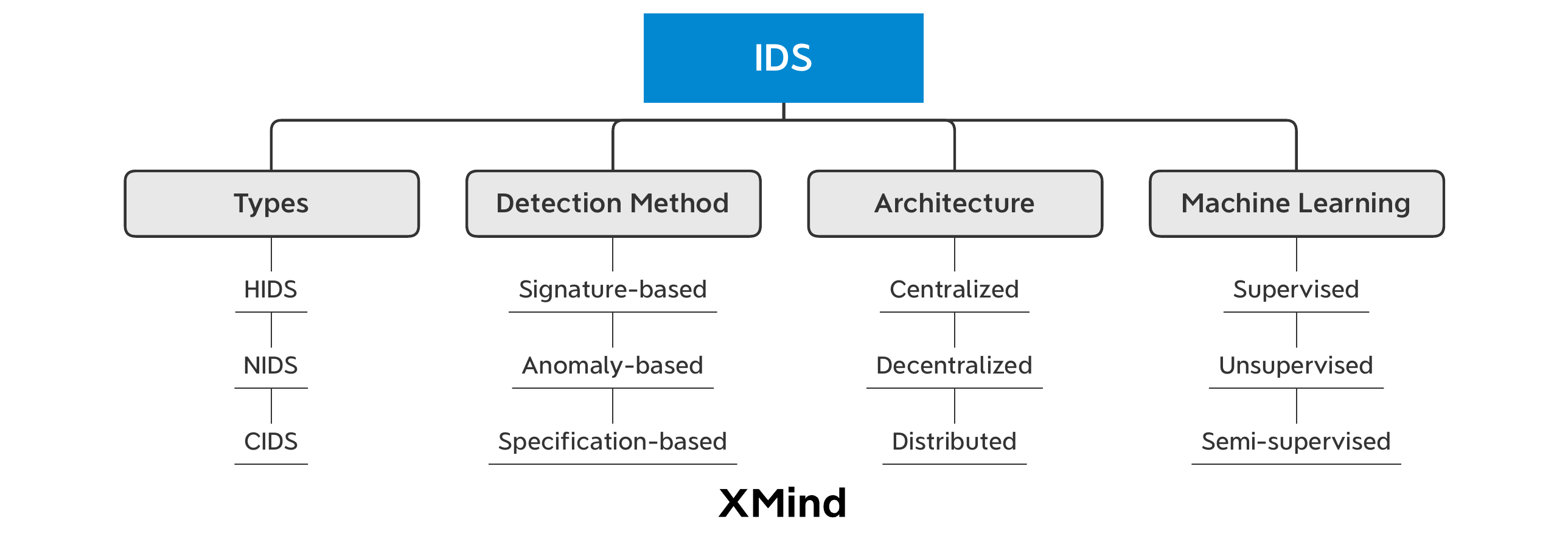}}
\caption{Taxonomy of IDS.}
\label{figure: Taxonomy of IDS}
\end{figure*}
\subsubsection{IDS Technologies}

IDSs can be classified in different ways. The taxonomy is illustrated in Fig.~\ref{figure: Taxonomy of IDS} and a comparison of each type of IDS is presented in TABLE~\ref{tab:IDSs}.

Many works have deployed IDSs in the smart grid context.
%
Lin et al. \cite{lin2016runtime} proposed a semantic analysis framework combining network-based IDS (NIDS) with power flow analysis to estimate the execution consequences of SCADA control commands. 
Firstly, the NIDS is specification-based and is implemented by Zeek (formerly named Bro) \cite{Zeek}, an open-source NIDS tool, to identify control commands on the SCADA network. Then, the extracted control commands are further processed by the power flow analysis software for predictive evaluation on their execution consequences. The results show that the semantic analysis framework achieved anomaly detection with 0.78\% FPR and 0.01\% FNR in 200ms for the large-scale 2736-bus system, satisfying the fast response requirement in SCADA systems.
Hong et al. \cite{hong2014integrated} targeted cyber threats against substations and proposed an anomaly-based collaborative IDS (CIDS) for IEC 61850-based substation automation systems. It includes both host-based IDS (HIDS) and NIDS to detect attacks targeting a single substation or simultaneously against multiple substations. It uses a WSU substation automation testbed to simulate simultaneous attacks against various substations. 
The proposed HIDS detects temporal anomalies according to system event logs generated from substation facilities like circuit breakers, IEDs, and user interfaces. An event log matrix is defined where each row represents same substation's anomaly indicators with consecutive time instants. Each column indicates host-based anomaly behavior in a certain type. By comparing consecutive row vectors, the proposed HIDS is able to detect temporal anomalies. It is further developed to discover simultaneous attacks among multiple substations by comparing the similarity of the event log matrix. 
Besides, the proposed NIDS detects malicious multicast messages, such as Sampled Measured Value (SMV) and GOOSE. The simultaneous intrusion detection method shows the ability to detect threats targeting multiple substations and identifying their locations. The proposed NIDS is a rule-based detection method by matching the multicast message packets with predefined rules. Thus, it is unable to detect unseen or unknown threats.

Besides, most of the recent IDS works are implemented with the assistant of ML due to its excellent performance in extracting features and detecting unseen or unknown cyberattacks.
Depending on whether the training data is labeled or not, ML-based IDS can be classified into supervised, unsupervised, and semi-supervised.
Supervised Learning trains detecting models with labeled dataset. It aims to find a map between inputs and their corresponding output. Representative approaches are classification and regression.
Unsupervised learning trains models with unlabeled data. It aims to find the similarity among input data and classify them according to the similarity distance. A typical unsupervised learning approach is clustering, where heterogeneous data are classified with metrics like Manhattan, Euclidean, and probabilistic distance \cite{tidjon2019intrusion,barlow1989unsupervised}.
Semi-supervised Learning trains detecting models using both labeled and unlabeled data.
He et al. \cite{he2017real} presented a deep learning based IDS to detect FDI attacks, especially those aiming at electricity theft, in real-time.
The proposed method contains a Deep-Learning Based Identification (DLBI) and a State Vector Estimator (SVE).
Firstly, the SVE detection scheme calculated the norm of measurement residual to detect bad data or maliciously injected data. Then, the rest of data will be processed by the DLBI for further evaluation. DLBI scheme uses Conditional Deep Belief Network (CDBN), which integrated the standard structure of deep belief networks with Conditional Gaussian-Bernoulli Restricted Boltzmann Machines (CGBRBM), to recognize the patterns of previous measurement data and use  extracted features to detect FDI in real-time.
The authors tested their detection scheme on IEEE 118-bus and IEEE 300-bus systems. They evaluated it with different previous-time observation window sizes, number of compromised measurements, number of hidden layers, environment noise levels, and the threshold value of SVE. Compared to SVM-based and ANN-based IDS, their scheme presented high detection accuracy even with occasional operation faults.
Also targeting energy theft behaviors, Yip et al.\cite{yip2017detection} proposed two linear regression-based algorithms to identify fraudulent customers and locate defective smart meters in the NAN. The authors presented that the energy consumption reported by smart meters on the consumer side should match the data recorded in collectors of electricity providers such as the substation. The deviation reveals the occurrence of energy theft or the existence of smart meters that are defective or compromised. As a result, the authors proposed two linear regression algorithms naming LR-ETDM and CVLR-ETDM, where CVLR-ETDM is an enhanced version due to the unstable performance of LR-ETDM in detecting inconsistent energy thefts. 
Faisal et al. \cite{faisal2014data} aim at detecting anomalies in the AMI networks. They regard network data as the stream because it is usually large, continuous, and fast transmitted in AMI networks. The authors chose seven classification algorithms of the Hoeffding tree in an open-source stream mining framework MOA \cite{bifet2010moa}. 
Because the smart meter, data concentrator, and headend vary in computing and memory resources, they evaluated them with accuracy and, specifically, the memory and time consumption. In addition, the authors comprehensively discussed the locations to deploy IDSs in the AMI. They proposed a less expensive way to install them in the existing AMI components.
In order to develop a context-adaptive and cost-effective IDS, Sethi et al. \cite{sethi2020context} proposed a hierarchical Deep Reinforcement Learning (DRL)-based IDS for accurate detection of novel and complicated cyberattacks. The proposed model consists of central IDS, distributed agent, state, action, and reward. The central IDS receives the actions made in agents and feedback reward to renew DRL model's parameters. When the agent's classification result is the same as the real result, the classifier will get a positive reward. The distributed agents are deployed in routers k-hop away from the central IDS and collect packets from corresponding end nodes. Data preprocessing and feature selection are executed in the agents, and the classification results are transmitted into the central IDS. To improve the robustness of their proposed model against adversarial samples, where attackers make small changes on the inputs to mislead the classification models, the authors tested their model with perturbations in their dataset and implemented a denoising autoencoder as a way to filter the perturbation of the data inputs. Their model was evaluated with three datasets, NSL-KDD \cite{Kdd}, UNSW-NB15 \cite{moustafa2015unsw}, and AWID \cite{kolias2015intrusion}. It presented an adaptive feature and robustness compared to other models.
Wang et al. \cite{wang2019detection} proposed the AdaBoost algorithm for multi-classification problems in detecting smart grid anomalies. It contains several classifiers which generate multiple predicted labels for the given input data. The classifiers are assigned with different weights according to the accuracy ratio on the training set of each classifier. In the end, the classifiers will vote for the final classification of the data. Feature construction and data processing were included in the training of the detection model and presented effectiveness in improving model accuracy. The model was trained and evaluated by an open-source PMU dataset in \cite{Powerdataset}. Compared to other ML algorithms, such as KNN, SVM, GBDT, XGBoost, and CNN, it achieved 93.91\% accuracy and 93.6\% detection rate higher than eight other prevalent techniques.
Otoum et al. \cite{otoum2019empowering} proposed a Reinforcement Learning-based IDS for wireless sensor networks (WSN). The presented Q-Learning-based model and their previously proposed Adaptive ML-based IDS were tested in a WSN with twenty sensors and evaluated on the KDDCup99 dataset \cite{tavallaee2009detailed}. They claimed that the Q-Learning-based model achieved almost 100\% success in detection, accuracy, and precision-recall rates, whereas the Adaptive ML-based IDS could achieve accuracy slightly above 99\%.
Al et al. \cite{al2018re} discussed the evaluation methods used in previous approaches and demonstrated that high accuracy could be reached for most of the ML models with properly tuned hyperparameters. In most research, the training and testing stages are processed on the same datasets sampled in the same environment having similar statistical distribution. The models evaluated by these strategies cannot truly reflect the practicality and performance in the real world. The models with high accuracy might be overfitting based on the specified dataset. It is hard to tell whether the model really learned the pattern of attacks or just represented a particular dataset. Therefore, the authors in \cite{al2018re} proposed an alternative evaluation strategy where different datasets with compatible sets of features are used for training and testing. The authors suggested to train and evaluate models using multiple datasets. It can be also used to assess the ability to detect zero-day attacks by checking whether the model can discover unseen attacks existing in another dataset.

\begin{table*}[tbp]
    \caption{Comparison of Different Types of IDSs}
    \label{tab:IDSs}
    \centerline{\includegraphics[scale=0.8]{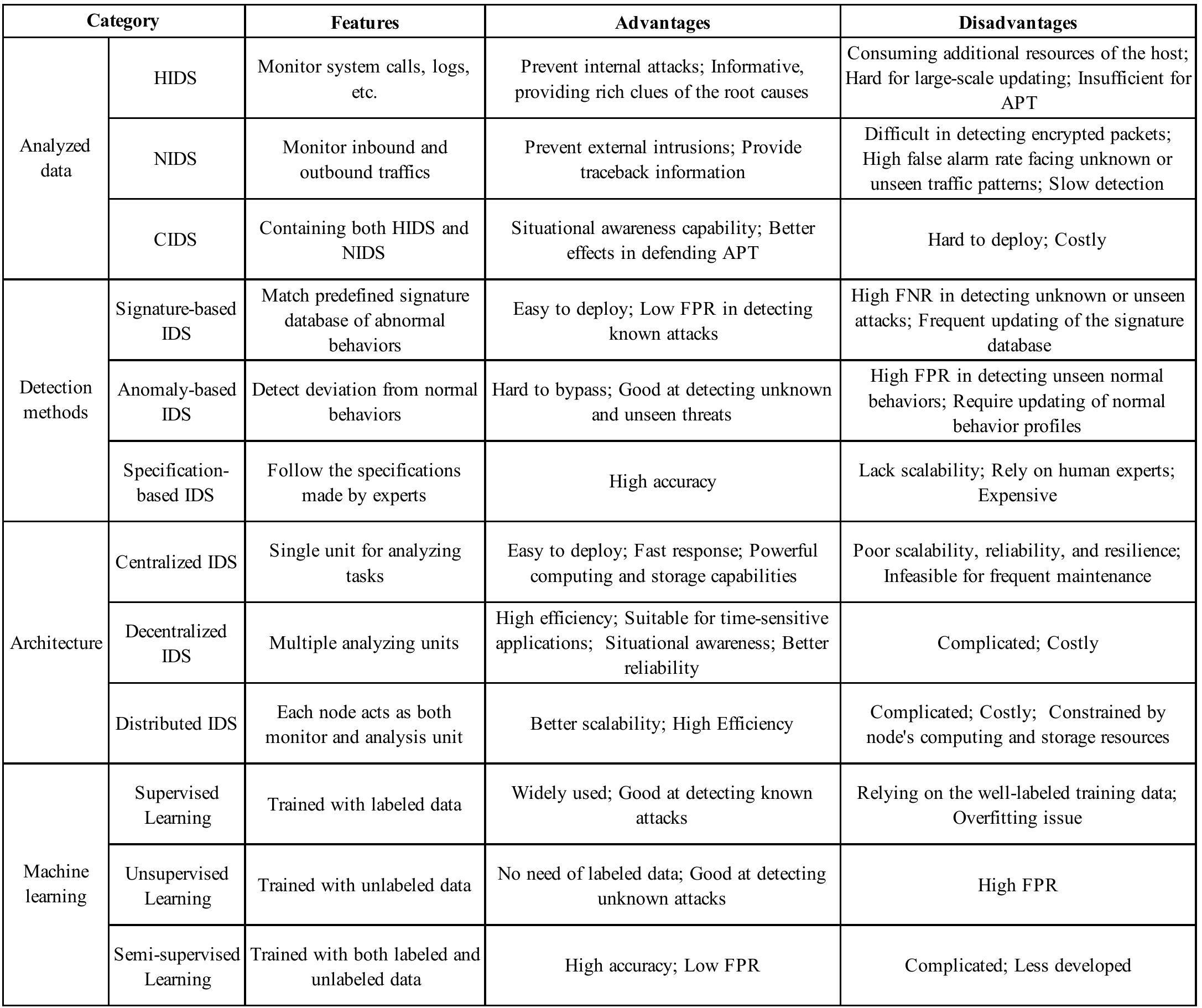}}
\end{table*}

\subsubsection{Challenges and Future Works}
The challenges and future works of applying IDSs in smart grid context are listed as follows.
\begin{itemize}
  
  \item \textit{Lack of Benchmark Datasets:}
  We notice that many researchers still use outdated open-source datasets (e.g., KDD cup 99 developed in 1999 and NSLKDD developed in 2009) to train their detection models. These datasets could no longer represent the features of novel cyberattacks. Researches established on these datasets may have academic meanings but are impractical to detect intrusions in real-world systems. Therefore, updated benchmark datasets are needed to cover features of prevalent attacks for supporting the development of novel IDSs and evaluating their performance. 

  \item \textit{Fast Detection in Time-sensitive Smart Grid Applications:}
  Smart grid applications could be time-sensitive. IDSs should satisfy the time constraints without adding large latency to the original communication channels. Therefore, IDSs have to work at near real-time either through enhancing the detection approaches or improving the computing and storage resources in the detecting points. 

  \item \textit{Trade-off between Accuracy and Complexity:}
  Resource-limited terminal devices and network nodes carrying detection tasks in the smart grid have decided the upbound complexity of intrusion detection algorithms. 
  Lightweight IDSs are needed as a trade-off between accuracy and resource consumption. 
  It is challenging because low accuracy means smart grid devices and networks can not be adequately protected, and the high FPR wastes human efforts for further confirmation. Therefore, resource consumption needs to be included in the design and evaluation phases of developing IDSs.

  \item \textit{Robustness and Security of the ML-based IDS:}
  Despite the advantages of ML-based detection approaches, the security vulnerabilities that existed in ML itself are also an important problem for recent researches. Firstly, ML algorithms highly depend on the quality of the training dataset. The models are built with the assumption that training and testing datasets have the same distribution \cite{papernot2018marauder}, which means that the data is usually sampled from the same network system environment. The model will have bad performance on the input data that does not belong to any classes of the training dataset. Moreover, the detection results of ML-based approaches depend on the input data and the parameters of detection models. It can be utilized by adversaries to evade detection systems by model inversion \cite{fredrikson2015model}, data poisoning \cite{biggio2012poisoning} or model evasion attacks \cite{biggio2013evasion}. Adversarial examples \cite{goodfellow2014explaining} can be artificially generated to misguide the prediction result of detection models. In addition, due to the vulnerabilities of ML models, valuable information about power systems, customers, and detection models can be intercepted or inferred by adversaries through model inversion and membership inference attacks \cite{shokri2017membership}.

\end{itemize}


\subsection{Honeypot}

To provide deep protection of the smart grid, the honeypot plays an important role in distracting adversaries' attentions, analyzing intrusion features, and capturing the evidence of illegal activities. It is designed as a vulnerable target that is attractive for intruders to compromise. It deceives attackers either with real devices or simulating their services and vulnerabilities to provide realistic responses \cite{mairh2011honeypot,dalamagkas2019survey}. Besides, honeypots keep interacting with adversaries to collect information for the analysis of attack features, tools, purposes, and strategies. It provides evidence of malicious activities intruding the smart grid systems and helps to establish the threat models so that defending approaches like the IDS could be updated based on them. Moreover, the honeypot consumes the time and computer resources of adversaries. The intrusion cost and risks of being captured will significantly increase. As a result, it is regarded as an active way to prevent attackers from penetrating smart grid systems and generate TI for long-term defense\cite{dalamagkas2019survey,wang2017strategic}.

This section introduces the honeypot and discusses its challenges and future works in the smart grid. 


\begin{table*}[htbp]
    \caption{Categories of the Honeypot}
    \label{tab:Categories of the Honeypots}
    \centerline{\includegraphics[scale=0.78]{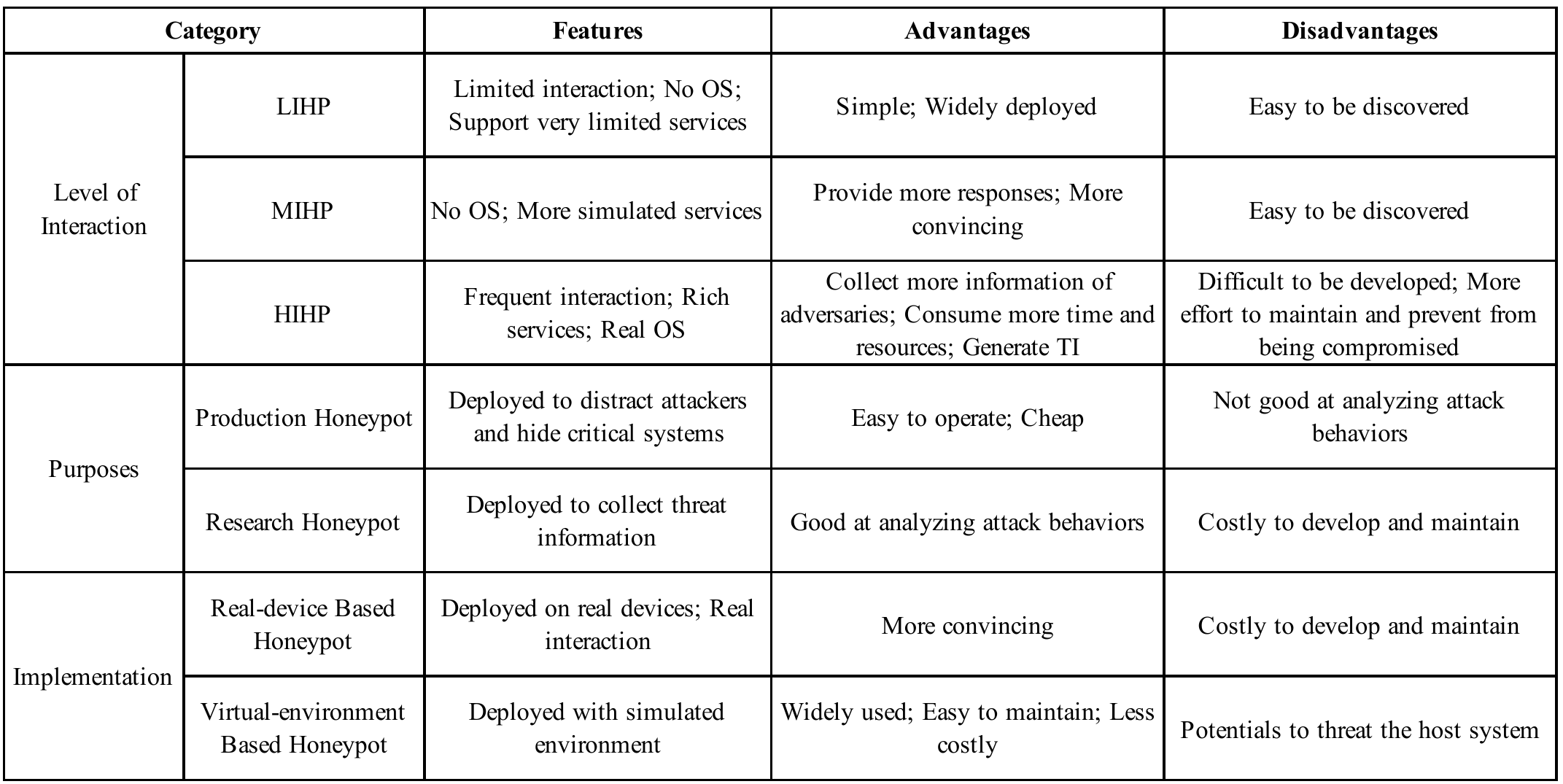}}
\end{table*}

\subsubsection{Honeypot Technologies}

Honeypots can be classified in different ways. The taxonomy and comparison of different types of honeypot is presented in TABLE~\ref{tab:Categories of the Honeypots}.

Many works have deployed honeypots in the smart grid context.
Conpot \cite{Conpot} is a widely used LIHP targeting ICS. It is easy to be deployed, modified, and extended. It could be expanded with novel configuration templates in the Extensible Markup Language (XML). It has been used to emulate ICS devices such as Kamstrup 382 smart meter and Siemens S7-200 PLC. It supports various protocols including TFTP/FTP, Modbus, SNMP, EtherNet/IP, IEC-104, and BACnet \cite{dalamagkas2019survey}. 
Jicha et al. \cite{jicha2016scada} analyzed and evaluated Conpot on AWS. They comprehensively described the experiment setup process of Conpot and the deployment of AWS. The authors deployed honeypots emulating Siemens SIMATIC S7-200 PLC, Guardian AST gas pump, IPMI, and Kampstrup smart meter devices. They evaluated Conpot with port scanners from Nmap and Shodan \cite{shodan}. The results proved that Conpot successfully depicted SCADA devices. However, the existence of additionally opened ports might expose the true identity of honeypots to malicious attackers.
%
Ferretti et al. \cite{ferretti2019characterizing} proposed a scalable network of LIHP based on the Conpot \cite{Conpot} framework for ICS. Integrated with an analysis pipeline, it achieved data enrichment, data analysis, and network traffic parsing. 
It captures ICS background noise traffic (e.g., traffic generated by network scanners) with no specific targets. The traffics could be generated during a multi-stage attack looking for potential victims. The authors deployed the proposed LIHPs in various network points to simulate ICS devices, such as PLCs, with representative protocols. They verified the proposed honeypot with Shodan \cite{shodan}, a searching engine for Internet-connected devices and proved it would not be easily identified as a honeypot by attackers. 
Besides, the parsing phase extracted information like source IP addresses, target ports, transport-layer protocols, and the application-layer request type and parameters from raw traffics.
It is enriched with external information like Autonomous System (AS) information, DNS PTR records, and country of source IP addresses to further analyze malicious behaviors. 
According to several months’ observation, all the background noise traffics in ICS are requests for device information or sensor readings.
Most of the traffics come from a few recurrent scanners, which are usually benign. However, there also exists a limited number of actors making different requests. The authors presented that identifying traffic actors is difficult since IP addresses are allocated to cloud hosting services or large ISPs.
%
%
%
Wang et al. \cite{wang2017strategic} proposed a Bayesian honeypot game model to address DDoS attacks in AMI networks. Considering that attackers can use anti-honeypots to collect information about security systems and bypass the defense mechanisms, the authors proposed honeypot game strategy and achieved the equilibrium condition between defenders and attackers using honeypots and anti-honeypots, respectively. According to the equilibrium, defenders can deploy honeypots more reasonably and achieve better effect on the protection of systems. The evaluation results based on constructed AMI network testbed proved that the proposed model reduced the energy consumption and improved the detection rate. 
%
Pa et al. \cite{pa2015iotpot} proposed a first IoT honeypot, named IoTPOT, and an analysis sandbox, called IoTBOX, 
aiming at attracting attacks leveraging Telnet to threaten IoT devices with various CPU architecture. 
In the proposed honeypot approach, the frontend low-interaction responder cooperated with backend high-interaction virtual environments to analyze the captured malware binaries. Based on the observation from IoTPOT and the analysis results from IoTBOX, four malware families spreading via Telnet were identified. The results proved that they are all used in DDoS attacks, and some malware families evolved and updated frequently, even in the limited observation period.
Previous works have demonstrated the effectiveness of honeypots in collecting valuable information about adversaries. The interaction processes and malicious behaviors are captured and stored in the log files for deep investigation. It decreases the false alarm rate of traditional IDS methods because all the interactions of honeypots can be considered malicious activities since no normal users will intentionally access the honeypot devices. As a result, it could also detect unseen attacks and zero-day attacks. The captured data can be used to recognize the pattern and attack tools, motivations, and strategies of novel adversaries and further support the making of defense strategies. Moreover, honeypots have a flexible structure where various honeypot software can be deployed in the network systems with different purposes. It can be combined with IDS or firewalls to provide prevention and detection functions.  More importantly, the honeypots actively distract attackers’ attention from critical system devices and have successfully increased adversaries' risks and costs by consuming their time and resources. It can be regarded as a deterrence mechanism to improve the security level of the smart grid.

\subsubsection{Challenges and Future Works}
The challenges and future works of exploiting honeypots in smart grid context are listed as follows.
\begin{itemize}
\item \textit{Risks of Being Compromised:} 
Honeypots that keep active interactions with adversaries have increased the risks of being compromised. Once it happens, honeypots could become slave bots controlled by attackers to threaten the security of smart grid systems. Therefore, the honeypots need to be well isolated from the smart grid once they are compromised. 

\item \textit{Easy to Be Discovered:}
Honeypots are only valuable when they can deceive adversaries to interact with them. 
However, honeypots, especially LIHP, simulating limited services and providing constrained responses are highly possible to be identified as a trap by experienced attackers. Then, they turn to be useless after exposing their true identities. Moreover, as mentioned in \cite{wang2017strategic}, anti-honeypot has been used to detect the existence of defending systems. Adversaries could send initiative packets to the target network to detect honeypot proxy servers. Once honeypot servers are detected, optional attack strategies will be made to bypass security systems and reach the real servers directly. 
As a result, the development of a high-fidelity honeypot that could deceive adversaries and decrease the possibilities of being identified will be a challenging work in future researches.

\end{itemize}


\subsection{Attribution}
\label{Attribution}
For smart grid intrusions, pure detection and blocking mechanisms are far from enough.
These methods only identify the existence of intrusion behaviors but cannot discover the actual adversaries behind the attacks, making them unavailable to solve security problems from the source. 

To provide deep protection of the smart grid, attribution becomes an indispensable part.
It has the ability to correlate distributed intrusion activities, summarize the characteristics of attack behaviors, infer attack paths, and figure out adversaries’ identities.
As a result, attribution is regarded as an active way to defend cyber threats against the smart grid by tracing adversaries' information.


\begin{figure}[bp]
\centerline{\includegraphics[scale=0.4]{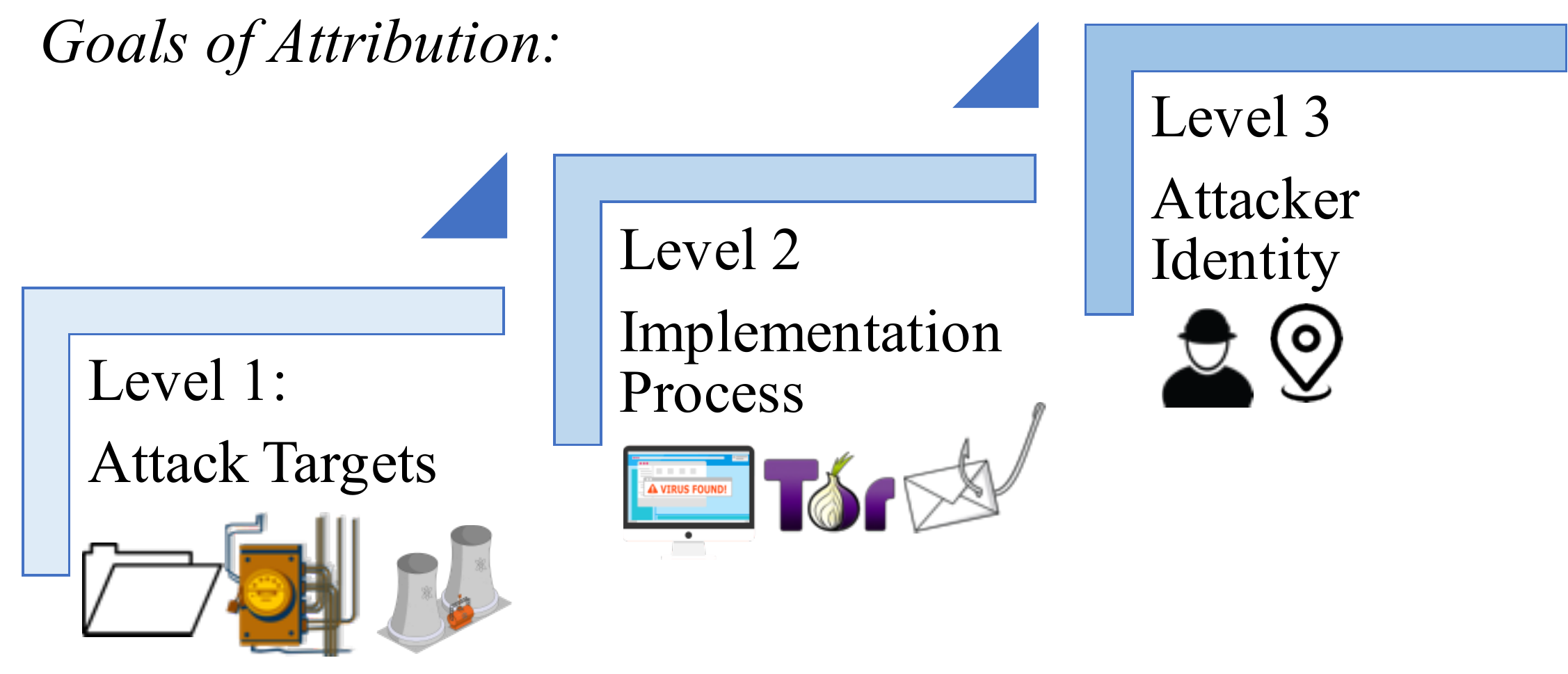}}
\caption{Goals of attribution in different levels.}
\label{Goals of Attribution}
\end{figure}

Besides, attribution is closely related to IDS and honeypots.
It uses the above techniques to collect adversaries’ information or detect malware, malicious users, stepping stones, and compromised servers. The difference is attribution aims to identify the attack purposes, attack processes, and information about adversaries’ true identities, as shown in Fig.~\ref{Goals of Attribution}. It includes the IP addresses, geographical locations, malware authorship, name of malicious organizations, the origin countries of targeting attacks, etc. Additionally, many techniques such as packet logging, packet marking, network flow watermarking, code authorship identification, and web tracing can be used for attribution or traceback. Attribution helps understand the attack strategies used by adversaries and consequently developing defending and responding strategies to improve the security level of the smart grid.

\subsubsection{Attribution Technologies}
In general, attribution can be realized in passive or active ways. 
The passive attribution monitors and analyzes the naturally exposed clues of malicious activities, such as system logs and network traffic, without interfering with the attack processes. Even it is practical and easy to be deployed in smart grid systems, the defenders cannot always capture the critical information required to identify illegal activities because sophisticated adversaries always try to hide attack traces and their identities using various techniques like code obfuscation, darknet, encryption, stepping stones, packet dropping, redundant packet padding, flow mixing, etc. 
Differently, active attributions leverage the interaction with adversaries to intentionally add watermarks on the traffic data or embed tracing programs to the files that adversaries could access to increase the chances of detecting adversaries. 
Here, we review the prevalent attribution approaches in code authorship identification, APT attribution, IP traceback, watermarking, and web tracking. 

\paragraph{Code Authorship Identification}
Code authorship identification aims at matching the stylometric features of programming code with their writers. Sophisticated adversaries could install malware into targeting smart grid systems for malicious activities. In order to actively infer the identities of attackers and provide forensic evidence to support criminal arrestment, code authorship identification will be an important step to reduce security threats from the root.
Abuhamad et al.\cite{abuhamad2018large} proposed "a deep learning-based code authorship identification system" (DL-CAIS) for large-scale, language-oblivious, and obfuscation-resilient code authorship attribution. The authors designed TF-IDF (Term Frequency-Inverse Document Frequency, a well-known tool for textual data analysis) based deep representation using multiple Recurrent Neural Network (RNN) layers and fully-connected layers for feature extraction. Then, a random forest classifier (RFC) was attached to process the extracted data and realize the large-scale code authorship identification. The authors evaluated their algorithm with the Google Code Jam (GCJ) dataset and code from 1987 public repositories on GitHub. It achieved a high accuracy (averagely over 94\%) under different experiment scenarios and proved effective while identifying authors using multiple programming languages and when the code was processed by obfuscation.  

\paragraph{APT Attribution}
APT is usually started by malicious organizations that have enough time, intelligence, and financial support. It could be a national behavior due to political or military purposes. Therefore, the inference of APT organizations is of great significance.
Rosenberg et al.\cite{rosenberg2017deepapt} proposed DeepAPT, a deep neural network (DNN) classifier, to identify the country responsible for APT behaviors. The authors used the sandbox to record the dynamical behaviors of APT malware. Further, the sandbox reports are regarded as raw feature inputs for the training of the DNN model. The authors presented that using raw features is cheaper and less time-consuming than the manual feature engineering process. It could prevent losing important information existing in the original data and achieve a higher accuracy using DNN. These are further demonstrated in their evaluation process. However, the authors also mentioned that APT organizations might subvert their proposed method to mislead the classification results and defame other irrelevant benign countries. Relative techniques, such as generative adversarial networks (GAN), can be used to modify their APT and cause misclassification of DNN models.

\paragraph{IP Traceback}
IP spoofing is generally used by malicious adversaries to hide their information from tracing and commonly exists in DDoS attacks. Yang et al.\cite{yang2012riht} reviewed the previous packet marking and packet logging based IP traceback schemes proposed in \cite{choi2004marking,malliga2008proposal,malliga2010hybrid}, which presented Huffman codes, Modulo/Reverse modulo Technique (MRT) and Modulo/REverse modulo (MORE), respectively, as the solutions of improving the usage efficiency of packet marking field, reducing the storage requirement for router logging and decreasing the number of routers required for logging. However, the authors in \cite{yang2012riht} proposed that previous methods still require high storage on logging routers. Their schemes might suffer from high FPR because of the collision of log tables and the loss of packet information when the routers refresh logged data. Moreover, the reconstruction of the packet transmitting path is also time-consuming due to the enormous amount of log data, especially under DDoS attacks. As a result, the authors proposed a novel hybrid IP traceback scheme RIHT, which marks routers’ interface numbers and integrates packet logging with a hash table. It only requires fixed storage in packet logging and does not need refreshing the logged tracking information. The evaluation presented zero FPR and FNR in route reconstruction and proved availability in detecting DoS attacks.

\paragraph{Watermarking}
\begin{figure}[tbp]
\centerline{\includegraphics[scale=0.75]{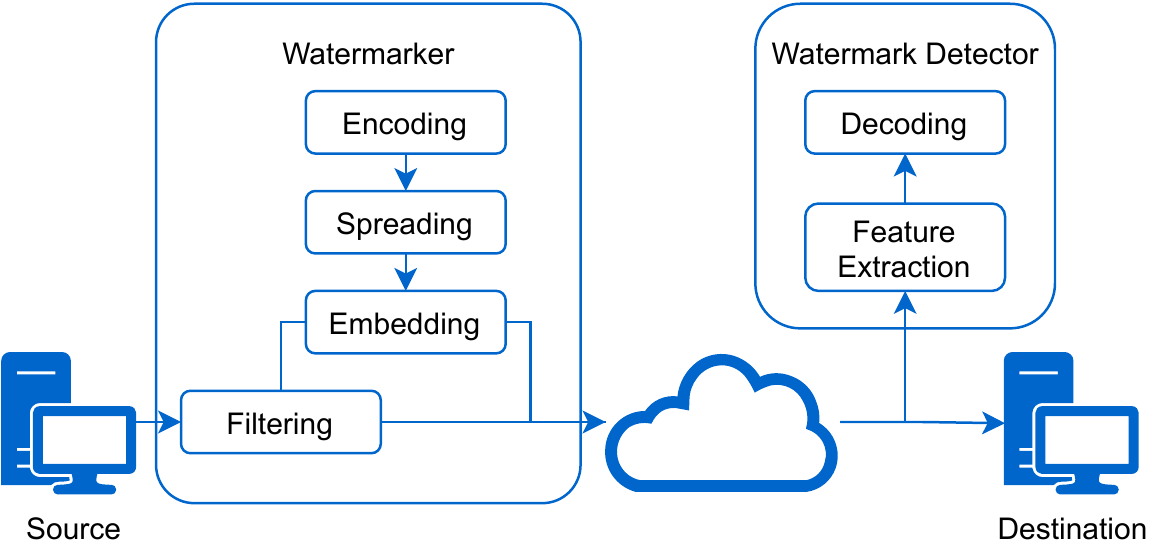}}
\caption{Architecture of watermarking \cite{iacovazzi2016network}.}
\label{watermarking}
\end{figure}
Network traffic watermarking is one of the active methods to trace malicious communications, especially of the botnet. Iacovazzi et al. \cite{iacovazzi2016network} reviewed the objectives, frameworks, and evaluation processes of watermarking and introduced the attacks against watermarking. 
Generally, watermarking embeds specific patterns to the selected traffics, which makes the traffics identifiable in other checking points. Fig.~\ref{watermarking} illustrates the work process of adding and observing watermarks in the traffic data. 
Watermarking can be used by defenders to identify a specific network flow, correlate different flows, reconstruct the transmitting route of malicious traffics and locate the positions of malicious users, such as botnet stepping stones and botnet masters.
Wang et al. \cite{wang2007network} proposed an interval centroid based watermarking scheme for low-latency anonymous communication systems. The conception of the interval centroid was introduced in their paper. By delaying packets within the selected flows, the interval centroid will change so that the difference can be further used as watermarks to link the senders and receivers. Based on their offline experiments and real-time experiments on anonymizing service provider www.anonymizer.com, the authors concluded that low-latency anonymous communications can be identified with the sufficiently long flow, even various flow transformations, such as adding cover traffic, packet dropping, and packet padding are used.
Houmansadr et al. \cite{houmansadr2009rainbow} proposed a timing-based watermarking algorithm named RAINBOW (Robust and Invisible Non-Blind Watermark). They recorded and correlated the timings of incoming and outgoing flows. Meanwhile, they produced watermark by delaying the packets with a small value. This value is chosen to be small enough which can be regarded as a normal transmitting jitter by malicious attackers to ensure the invisibility of the proposed algorithm. 
Further, the authors proposed an interval-based watermark SWIRL (Scalable Watermark that is Invisible and Resilient to packet Losses) in \cite{houmansadr2011swirl} which is considered as the first watermark that is practical for large-scale traffic analysis. Different with other interval-based watermarking, SWIRL introduces little delays to the network flow. SWIRL produces watermarks depending on the characteristics of selected flows. The proposed algorithm was evaluated on PlanetLab testbed \cite{bavier2004operating} and presented very low false error rates on short flows. The authors presented that SWIRL can be used for resisting multi-flow attacks, detecting stepping stones, linking anonymous communications and detecting congestion attacks on TOR.
Iacovazzi et al. \cite{iacovazzi2017dropwat} proposed a timing-based watermarking algorithm, DropWat, for tracing data exfiltration attacks aiming at stealing sensitive information from a private network to unauthorized servers. DropWat indirectly modifies the interpacket delays of selected packets by dropping packets pseudo-randomly. The authors claimed that DropWat watermark is invisible to attackers due to the same behaving patterns between natural packet loss events and intentional packet dropping. DropWat was tested in the scenarios that data exfiltration attackers use stepping stones or TOR (The Onion Router) anonymous networks to hide their identities. It was implemented with the assumption that stepping stones will not propagate packet losses, for example using TCP protocols. The evaluation results proved that DropWat is effective for data exfiltration traceback even with a high transfer rate. Additionally, it is effective for attacks using TOR networks. However, because the packet dropping behaviors should not affect the normal throughput of original traffics, DropWat is not suitable for short-lived or interactive flows.

\begin{figure}[bp]
\centerline{\includegraphics[scale=0.55]{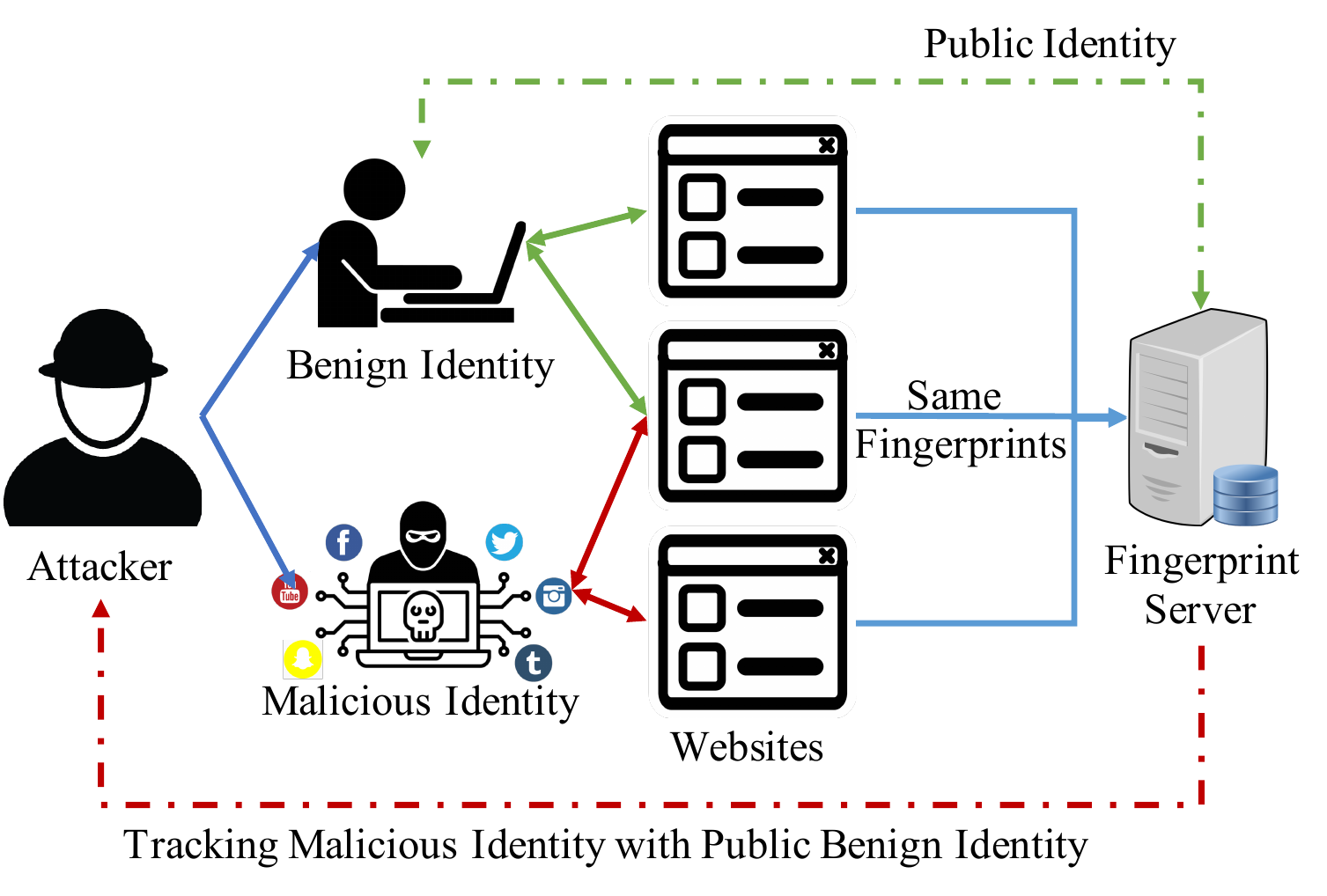}}
\caption{Attribution based on browser fingerprinting in crossing websites scene.}
\label{Browser fingerprinting}
\end{figure}

\paragraph{Web Tracking}
Web tracking recognizes and correlates the identities of past website visitors for user authentication, identification, or delivering personalized services, such as business advertisements. The significance of web tracking in smart grid scenarios stays in two aspects. On the one hand, cyber attackers could use social engineering methods to collect the personal information of staff working in the smart grid enterprises through web browsers. On the other hand, attacking tools, such as malware, could be delivered to the smart grid facilities through malicious or phishing websites. Besides, attackers usually own both benign public identities for normal daily network activities and hidden malicious identities for illegal activities at the same time. The behavior patterns of benign identities can be extracted and further utilized to correlate and identify the hidden malicious identities of the same attackers (shown in Fig.~\ref{Browser fingerprinting}). Different browser fingerprinting techniques have been developed to identify attackers’ activities on the same browsers, multiple browsers, or even browsers on different devices.
Cao et al. \cite{cao2017cross} proposed a (cross-) browser fingerprinting technique which is the first to use novel OS and hardware level features, e.g., those from graphic card, CPU, audio stack and installed writing scripts, for both single and cross browser fingerprinting. They provided open-source implementation on Github. The evaluation results presented that their approach is able to fingerprint 99.24\% of users on a single browser.
Starov et al. \cite{starov2017xhound} focused on the fingerprinting of browser extensions, e.g., AdBlock and Ghostery, and proposed Xhound (Extension Hound) as the first fully automated system for fingerprinting browser extensions. Xhound fingerprints the side effect of browser extensions on a page’s DOM, such as adding new DOM elements and removing existing ones. The authors’ research proved that browser extensions are fingerprintable and offered a new methodology for web tracking techniques.

\subsubsection{Challenges and Future Works}
The challenges and future works of attribution approaches are listed as follows.
\begin{itemize}
  \item \textit{Robustness of Code Authorship Identification:}
  The features of code writers vary from factors like programming habits, education levels, languages, and programming platforms. However, larger-scale programs usually involve multiple writers. It challenges the robustness of identification tools. Besides, novel malware could evolve from other tools, which means developers can copy other authors’ code to develop their own attack tools. This makes the authorship identification results unreliable. 
  Moreover, adversaries can intentionally add other organizations' features to their tools to misguide defenders and set up other countries for political and military purposes. The inference results of most researches can only be regarded as a reference instead of confirmed evidence. Therefore, future works need to solve the problems in identifying programs with multiple writers and enhance the robustness to prevent being misled by adversaries.

  \item \textit{Attribution Overheads:}
  Passive traffic analysis methods and network flow watermarking technologies could introduce an extra burden to the smart grid systems. It either requires extra storage and computing resources for routers to log and search information of passed flows or delays the normal communication process by adding a timing-based watermark to the original flows. However, smart grid systems could have strict limitations on resources and communication latency. Therefore, exploiting attribution approaches needs to specifically pay attention to the requirements of the smart grid. 
  \item \textit{Robustness and Concealment of Watermarking:}
  Firstly, manually injected watermarks can be interfered by natural noises in the communication channels. Secondly, adversaries might discover the injected watermarks and bypass watermarking approaches by intentionally dropping packages or adding meaningless redundant packets. As a result, robustness and concealment of watermarking should be paid more attention to mitigate the interference from both natural noise and attackers.
\end{itemize}

\subsection{Threat Intelligence}

Threats targeting the smart grid are getting complicated and turn to be distributed and delivered in multiple stages.
Therefore, the defending mechanisms have to be aware of the threat situation in the global systems to extract attack features and update defending strategies to enhance the security abilities.
In order to do so, the threat intelligence (TI) becomes vital important. 
According to the definition given by Gartner, the TI is regarded as “evidence-based knowledge, including context, mechanisms, indicators, implications and actionable advice, about an existing or emerging menace or hazard to assets that can be used to inform decisions regarding the subject's response to that menace or hazard\cite{TheatIntelligence}.”
The significance of the TI is to collect threat information from various sources, extract and summarize into different forms, and share them with protected systems to achieve real-time situation awareness, enhance the fast responding abilities, and complete defending mechanisms to prevent being compromised by the security threats.

\subsubsection{TI Technologies}
According to Tounsi et al. \cite{tounsi2018survey}, the TI can be generally classified into long-term strategic TI, tactical TI, and short-term operational TI, technical TI.

\begin{itemize}
  \item \textit{Strategic TI:}
  Strategic TI aims to assist analysts in judging the current and upcoming threat situations to make defense strategies and response decisions. Strategic TI is usually in forms of reports, briefings, or conversations \cite{tounsi2018survey} which includes the risk assessment, financial impact, and situation prediction to support security strategies.

  \item \textit{Tactical TI:}
  Tactical TI helps defenders learn adversaries’ tactics like attack tools to improve the resiliency and security of protected systems. It can be gathered from public shared technical articles in TI sharing communities or TI service providers.

  \item \textit{Operational TI:}
  Operational TI provides more specific and detailed information about attacks, such as malware code. This type of TI is more valuable but challenging to obtain. It can usually be accessed when the organizations are compromised or gathered from open-source intelligence providers and private forums.

  \item \textit{Technical TI:}
  Technical TI is usually applied to the defense mechanisms such as IDS, firewalls, etc. It provides indicators of compromise (IOC) to defense mechanisms so the defense tools can be updated and upgraded to detect and prevent corresponding attacks.
\end{itemize}
According to our observation, the top three topics about TI are its collection, sharing, and analysis phases. Therefore, this section reviews the critical works in the three phases and proposes its challenges and future works.


\paragraph{TI collection}
TI or evidence-based knowledge is collected from multiple sources. It can be the data (e.g., bot IP addresses, domain names or file hashes \cite{li2019reading}, malware signatures, network traffic, etc.) collected from security mechanisms such as IDS, honeypot, malware binaries analysis, system logs, and attribution techniques. These security technologies extract IOC about adversaries to update detection rules and mitigation methods to improve the accuracy of protection and provide proper remediation strategies as a reaction to inferred attacks.
Bou et al.\cite{bou2017cyber} proposed a novel approach that exploits cyber TI to identify CPS attacks, trigger remediation strategies in the physical realm, and build countermeasures to provide CPS better resiliency. The proposed approach consists of cyber layer dynamic malware binary analysis, honeypot, physical layer CPS monitor, and a cyber-physical threat detector. In the cyber layer, dynamic malware binary analysis generates cyber TI by running malware samples to produce an XML report describing malware activities. Additionally, the authors implemented a Conpot-based honeypot to provide external TI. Utilizing the signatures generated by malware analysis, honeypot, and the cyber-physical data flows collected from CPS monitors, the authors built a cyber-physical threat detector with semantic graphs. Then, it compares the similarity among the semantic graphs to identify suspicious threats in cyber-physical data flows extracted from the communication channels.
Moustafa et al. \cite{moustafa2018new} proposed a TI scheme for CPS and industry 4.0 systems. The proposed scheme contains a smart data management module and a TI module. The smart data management module processes heterogeneous data collected from sensors, actuators, and network traffic. The TI module is based on Beta Mixture-Hidden Markov Models (MHMMs) to identify malicious activities in both physical and cyber domains. Due to the lack of available public CPS datasets, the authors aggregated both power system dataset of sensors and physical devices \cite{Powerdataset} and UNSWNB15 dataset of network traffic \cite{moustafa2015unsw} to form a complete CP dataset for the training and validating of the TI module.
In addition to collecting TI through classical security mechanisms, the TI shared among public and private online communities, forums, and social communicating platforms, especially through the darknet, has successfully attracted researchers’ attention. Valuable information about system vulnerabilities or even open-source attack tools can be found trading and exchanging on these platforms. Adversaries can directly utilize the exposed vulnerabilities to build their own attack tools or directly leverage TI or attack tools selling on the darknet to achieve their goals. On the other hand, it also makes it a vital important channel to obtain TI.
Nunes et al. \cite{nunes2016darknet} 
proposed a system to gather TI from marketplaces providing attack tools and online forums in the darknet or deepnet.
The authors presented that the proposed system collected on average 305 high-quality cyber threat warnings each week, including novel malware that has not yet been applied.
Li et al. \cite{li2019reading} focus on the assessment and evaluation of shared TI data. They introduced several public and private intelligence providing platforms, including Facebook ThreatExchange, Paid Feed Aggregator, Paid IP Reputation Services and public blacklists and reputation feeds such as Badips and Packetmail. Among them, Packetmail IPs and Paid IP Reputation captured threat data via security mechanisms like honeypots, analyzing malware, etc. Differently, Badips and the Facebook ThreatExchange obtained data from general users or organizations who have been attacked and are willing to submit the indicators (e.g., malicious IP address and file hash) to these TI providers.
Samtani et al. \cite{samtani2016azsecure} presented that there were many cyber TI portals and malware analysis portals that passively obtained TI data from anti-virus engines, network traffic data, IDS, IPS, event logs, and malware binaries analysis. They highly rely on the community submissions of malware. However, none of these TI portals and malware analysis portals collect data directly from hacker communities. Therefore, the authors designed a novel TI and malware portal, AZSecure Hacker Assets Portal, to collect data from hacker communities. Moreover, ML techniques were applied to collect and analyze the hacker source code and attachment assets from hacker forums.

\begin{figure}[tbp]
\centerline{\includegraphics[scale=0.4]{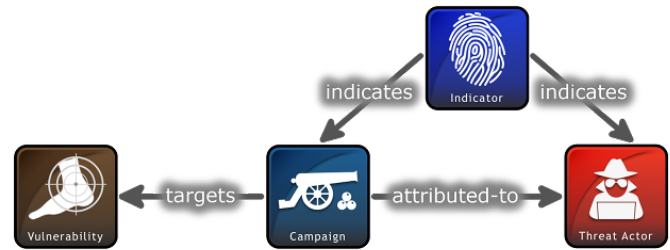}}
\caption{STIX 2 relationship example \cite{oasis2019introduction}.}
\label{STIX 2}
\end{figure}

\paragraph{TI Sharing}
Sophisticated adversaries could attack smart grid systems with multiple stages and target distributed smart grid components to expand the threat impact to trigger cascading failures and cause large-scale blackouts.
Consequently, security mechanisms only depending on the local TI to update is difficult to cope with the evolving and rapidly occurring cyberattacks. 
Therefore, the sharing of TI is of great significance for enhancing the security level of the power systems.
Collected TI is encouraged to be shared among internal institutions and different organizations across enterprises and countries. 
At the same time, a unified TI representation and sharing standard should be established to facilitate better understanding and communication between different organizations.
Currently, Structured Threat Information eXpression (STIX) has become one of the most widely used standards for the representation of TI. It has specified a language and serialization format for TI sharing \cite{tounsi2018survey,barnum2012standardizing}. 
The updated version STIX 2.1 consists of 18 objects: attack pattern, course of action, campaign, grouping, indicator, identity, intrusion set, infrastructure, location, note, malware, malware analysis, observed data, report, opinion, tool, threat actor, and vulnerability. 
By connecting multiple objects through their relationships, TI can be clearly represented by STIX (as illustrated in Fig.~\ref{STIX 2}\cite{oasis2019introduction}). 
STIX integrates many standards to provide diversity and more available functions. To support the sharing of TI represented in STIX, Trusted Automated Exchange of Intelligence Information (TAXII) is design as an application protocol for TI sharing over HTTPS \cite{oasis2018introduction}. Besides, TAXII also supports TI sharing in other formats besides STIX. Moreover, Cyber Observable eXpression (CybOX) is commonly used in STIX as a structured language to represent cyber observables, which can be utilized for TI, logging, malware characterization, indicator sharing, incident response, digital forensics \cite{Cybox} etc.
\paragraph{TI Analytics}
Qamar et al. \cite{qamar2017data} proposed that the TI sharing reports are usually generated manually by security professionals, which could contain incomplete and incorrect information. Some of the reports even cover repeated information about existed attacks. In addition, large volumes of intelligence data are included in TI containing redundant data to make the usage of TI inefficient. Lack of assessment and validating of provided TI data, TI analytics are desired to analyze and extract valuable information from gathered TI. It is also important to relate the TI information with specific network configuration to find the meaningful TI satisfying the requirement of specific scenarios.
Consequently, Qamar et al. \cite{qamar2017data} proposed a TI analytics framework applying Web Ontology Language (OWL)-based ontology as a semantic model to represent CybOX, STIX, network configurations, and CVE information to relate large volumes of shared TI information with network configurations to identify network associated threats, find out their relevance, and evaluate the risk on the network.
Liao et al. \cite{liao2016acing} focused on IOC published in TI articles and proposed IOC Automatic Extractor (iACE) to automatically discover helpful TI information and relate the IOC token (e.g., zip file) and its context (e.g., “download,” “malware”) within the text of TI articles. 
The proposed scheme applied graph mining techniques to discover tokens that are related following the common IOC description ways. iACE will further extract these tokens and transform them into a standard IOC format, OpenIOC, including both indicators and their context.
Li et al. \cite{li2019reading} presented the problem of lacking assessment of TI data, which makes it difficult for TI product consumers to compare various TI sources and choose the most valuable ones. Therefore, the authors specified six TI metrics, including volume, differential contribution, exclusive contribution, latency, accuracy, and coverage, and successfully utilized them to analyze and evaluated different TI sources.

\subsubsection{Challenges and Future Works}
The challenges and future works of TI is listed as follows.
\begin{itemize}
  \item \textit{Quality of TI:}
  A significant amount of TI data is generated and collected by each organization using different security tools. However, the practicality and quality of generated TI data are uncertain due to the lack of standardized assessment and evaluation methods. Manually generated TI reports could include incorrect and incomplete TI information. In addition, redundancy existing in a large amount of TI information needs automated, intelligent analytics methods to filter, extract, and relate relative information for practical usage. 
  However, it is hard for small institutions to process their generated data and provide precise labels or separate different categories of threats. 
  As a result, it provides great market opportunities for professional TI analyzing organizations to sell timely TI information to smart grid consumers to enhance smart grid security mechanisms and protect power systems from zero-day attacks and other sophisticated attacks.

  \item \textit{Obstacles in TI Sharing:}
  TI sharing is able to enhance the security architecture by collaboratively learning the threat trends and system situations to reduce the risks of compromising system units or preventing cascading events. The experience from systems that have suffered CPS threats can be borrowed to complete and enhance the defense mechanisms of their own systems. However, there are still several obstacles stopping organizations from confidently sharing TI information. Firstly, exposing vulnerabilities of a product could make the producers lose advantages in the market and affect consumers' purchasing choices. Secondly, the shared TI might be utilized by untrusted participants for malicious or competitive purposes. Additionally, the sharing of TI is usually accompanied by extra costs and an increasing budget. When the benefits of TI providers are at the risk of being hurt, or there are no economic returns for sharing TI, organizations will not be willing to participate in TI sharing activities. Moreover, the TI sharing process is quite subjective for security managers. Analysts will choose not to share their gathered information if they think it is less valuable or unaware of the intrusion of malicious attacks.
\end{itemize}

As a result, future efforts on TI should be made on improving the quality and timeliness of TI, encouraging TI sharing and overcome the obstacles of TI sharing.

\section{Digital Twin in the smart grid}

\label{DT in SG}
































  DT is a novel technology attracting the attentions of smart grid experts. 
  From the perspective of a CPS, DT acts as a paralleled digital representation of real-world cyber-physical entities, including physical devices, systems, network elements, etc. 
  Since the smart grid is a typical CPS, the development of DT in smart grid context becomes vital important.
  DTs in the smart grid should share exactly the same or part of the critical features of original real-world power grid entities (including electricity entities and network elements) so that additional functions can be realized based on them. These functions could be visual display, real-time surveillance, failure diagnosis, security test, risk assessment, state prediction etc.

\begin{figure*}[htbp]
\centerline{\includegraphics[scale=0.9]{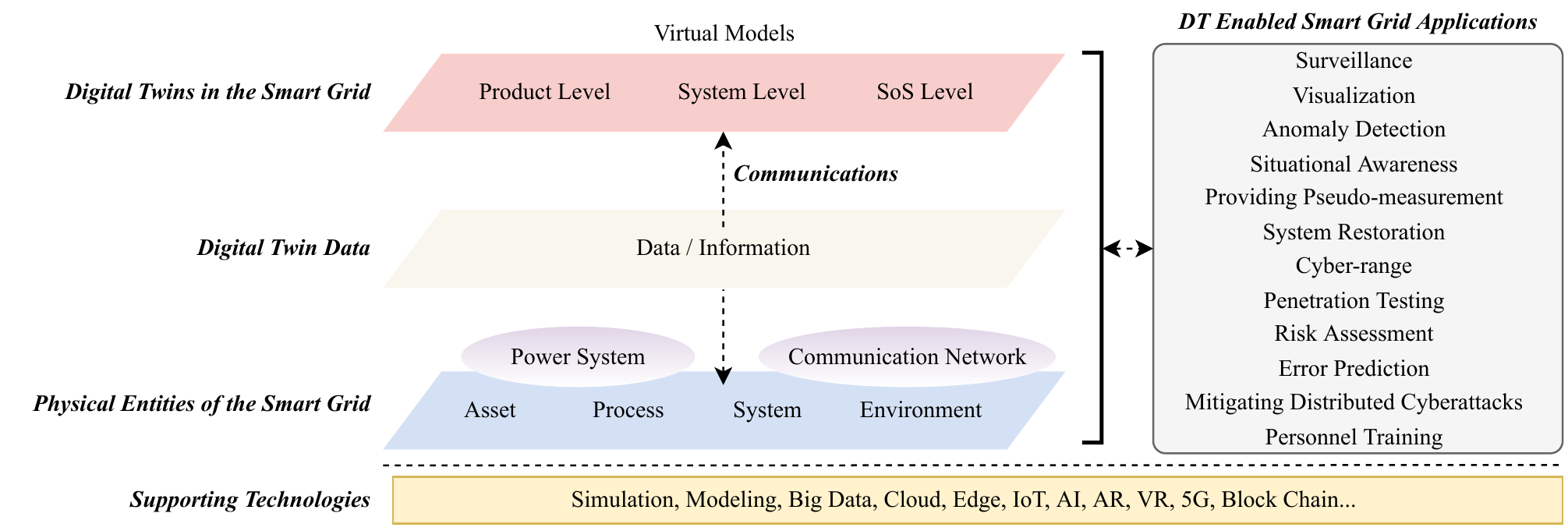}}
\caption{Digital twin's structure and services in the smart grid context.}
\label{fig: A Scalable Digital Twin Architecture}
\end{figure*}

  This section introduces the concept and development of DT, reviews the applications of DT in the smart grid, and indicate DT as an enabler for enhanced cybersecurity. A summary of the DT structure and enabled applications is illustrated in Fig.~\ref{fig: A Scalable Digital Twin Architecture}.

\subsection{Digital Twin}
\label{sub_DT}


DT is a virtual representation of a real-world object, system, or process with real-time updated data\cite{GartnerDTdefine,mikell2018cheat}. 
It is initially proposed by Michael Grieves in 2003 for his University of Michigan Executive Course on PLM in industrial manufacturing context\cite{grieves2014digital}.
In \cite{grieves2014digital}, Grieves indicated that a DT should consist of three parts, including physical entities in the real space, virtual models in the virtual space, and data or information connection between them. 
The physical part collects measurement data to learn product states or the performed operations.
It is reflected to the virtual space through 3-D modeling so that DT could visualize the product, monitor products states, and guide manufacturing processes.
%
%
%
Tao et al. \cite{tao2018digital} modified Grieves' work and proposed a DT-driven PHM method for complex equipment with a five-dimension model, including physical entities, DT data, virtual equipment, services, and the connection among them. They emphasize the DT data as a data integration of more comprehensive and accurate information originating from various sources (e.g., domain knowledge, services, physical entity, virtual equipment) to provide services like power output monitoring. 
In their other works, they designed a DT-based shop-floor \cite{tao2017digital}, discussed the product design, manufacturing and services driven by DT \cite{tao2018design}, and summarized the existing works about DT in \cite{tao2018digitaltwin}.


In the mobile communication context, many works \cite{liu2020vision,dong2019deep,sun2020reducing,dai2020deep,tao2021digital} have proposed the "DT network" as a real-time mirroring of physical networks, where management and control strategies (e.g., resource allocation, energy-saving, computation offloading) can be made, optimized, tested, and validated in the DT model before applied to the realistic network.
Besides, due to the service level agreement (SLA) requirements of 5G network, Sun et al.~\cite{sun2021digital} applied DT for network operation and maintenance.
Network service information, including service rules, network configuration parameters, resource information, and operation conditions, will be synchronized to the DT network to evaluate whether it satisfies the SLA. Based on the analysis of AI, DT network could provide intelligent optimization suggestions and verify them in the DT network.


In summary, the object of DT could be a physical product, process, system, and network. 
From the perspective of CPS, DT can be a paralleled digital representation of any real-world cyber-physical entities. The DT model could be a encapsulated software or model stored in cloud or edge servers and every single model could connect with each others to construct a giant simulated system or network. 
In general, these features enabled DT to provide visualization and monitoring of the physical entity, and promote the design, management, and adjustment processes. 
All the optimization strategies could be made based on the analyzing of DT collected information with the assistant of AI algorithms. 
Then, these strategies can be firstly evaluated and tested before deploying into the physical entities.
From another point of view, these features provide DT the opportunities to be an enabler for enhanced security, especially cybersecurity, due to its potentials in monitoring system states, detecting anomalies through AI-based analysis of DT data, predicting future situation, supporting decision making, and evaluating security strategies in the DT environment.


\begin{table*}[htbp]
\centering
\caption{SUMMARY OF DIGITAL TWIN RESEARCHES}
\label{tab:DT works}
\begin{tabular}{|c|c|c|}
\hline
\textbf{Topics} &
  \textbf{Contents} &
  \textbf{References} \\ \hline
DT Concept &
  DT   is a digital representation of real-world cyber-physical entities with  syncrhonized states &
  \cite{GartnerDTdefine,mikell2018cheat} \\ \hline
Basic Components &
  Physical   entities, Virutal models, DT data, Data Interaction, DT applications &
  \cite{grieves2014digital,tao2018digital,jiang2021novel} \\ \hline
DT Object &
  Physical   product, network element, process, system, environment &
  \cite{GartnerDTdefine,mikell2018cheat,liu2020vision,dong2019deep,sun2020reducing,dai2020deep,tao2021digital,sun2021digital} \\ \hline
\multirow{3}{*}{DT Fidelity} &
  Unit   Level &
  \cite{jiang2021novel,moutis2020digital,zhou2019digital} \\ \cline{2-3} 
 &
  System Level &
  \cite{darbali2021state,he2018surveillance} \\ \cline{2-3} 
 &
  SoS Level &
  \cite{saad2020iot,brosinsky2018recent} \\ \hline
\multirow{12}{*}{DT-enabled Functions} &
  Surveillance &
  \cite{he2018surveillance} \\ \cline{2-3} 
 &
  Visualization &
  \cite{he2018surveillance} \\ \cline{2-3} 
 &
  Anomaly/Intrusion Detection &
  \cite{zhou2019digital,brosinsky2018recent,he2018surveillance,eckhart2018specification} \\ \cline{2-3} 
 &
  Situational Awareness &
  \cite{brosinsky2018recent,salvi2022cyber} \\ \cline{2-3} 
 &
  Providing Pseudo-measurement &
  \cite{darbali2021state,moutis2020digital} \\ \cline{2-3} 
 &
  System Restoration &
  \cite{saad2020iot} \\ \cline{2-3} 
 &
  Cyber-range &
  \cite{becue2018cyberfactory,salvi2022cyber} \\ \cline{2-3} 
 &
  Penetration Testing &
  \cite{atalay2020digital,bitton2018deriving} \\ \cline{2-3} 
 &
  Risk Assessment &
  \cite{atalay2020digital,hadar2020cyber} \\ \cline{2-3} 
 &
  Error Prediction &
  \cite{he2018surveillance} \\ \cline{2-3} 
 &
  Mitigating Distributed Cyberattacks &
  \cite{saad2020implementation} \\ \cline{2-3} 
 &
  Personnel Training &
  \cite{brosinsky2018recent,becue2018cyberfactory,salvi2022cyber} \\ \hline
\multirow{4}{*}{Field} &
  Design,   Production, and Manufacturing &
  \cite{grieves2014digital,tao2018digital,tao2017digital,tao2018design,tao2018digitaltwin} \\ \cline{2-3} 
 &
  Communication Networks &
  \cite{liu2020vision,dong2019deep,sun2020reducing,dai2020deep,tao2021digital,sun2021digital} \\ \cline{2-3} 
 &
  Smart Grid &
  \cite{jiang2021novel,moutis2020digital,darbali2021state,zhou2019digital,he2018surveillance,saad2020iot,brosinsky2018recent} \\ \cline{2-3} 
 &
  Cybersecurity &
  \cite{becue2018cyberfactory,ITEACyberFactory,atalay2020digital,saad2020implementation,bitton2018deriving,hadar2020cyber,salvi2022cyber,eckhart2018specification,gehrmann2019digital} \\ \hline
\end{tabular}
\end{table*}


\subsection{Digital Twin in the Smart Grid}
\label{sub_DT in SG}


Similar with the concept of CPS, DT realizes the convergence of physical and cyber domains.
Some researchers regard it as the next phase of the CPS \cite{kan2019digital}. 
Since the smart grid is a typical CPS covering both industrial systems and communication networks, the functions of DT should also be applicable for the smart grid context.




%
The construction of DT establishes a platform for intelligent smart grid applications with unified information source. Upper smart grid management businesses could directly connected to the DT to save resources and provide better security.
However, the lack of a standard DT model makes the practical construction of DT for physical products hard to implement. 
Therefore, Jiang et al. \cite{jiang2021novel} proposed a four-dimensional DT model named "OKDD" to guide the digitization of smart grid physical entities. It consists of the ontology-body, knowledge-body, digital-portal, and data-body. Firstly, the ontology-body includes the standardized description and identification of physical entities in the smart grid. Secondly, the knowledge-body is an integration of knowledge stored in various theoretical models, e,g., electromagnetic, temperature, motion, etc. Thirdly, the data-body organizes and manages the data generated during physical entity's entire life cycle to improve the efficiency of extracting the desired information. Then, the digital-portal provides secure information interaction for data services. 
Based on the model, the authors constructed the DT model for the vacuum circuit breaker and extended it to a 35 kV substation. As well, they developed a PHM system based on the DT model in a 110 kV substation.



Moutis et al. \cite{moutis2020digital} proposed a unit-level DT model for the distribution power transformer. It realizes the real-time monitoring of medium voltage (MV) waveforms of voltage, current, and active and reactive power without MV instrumentation but using the DT model and low voltage (LV) side measurements to calculate MV. The calculated MV waveforms through transformer DT models are as accurate as the instrument measured data. It is easy to be deployed, more economic efficient and has avoid the disruptions to power grid operations.


%
%
The introduction of renewable energy has brought unstable factors, like voltage swings, to the smart grid.
Therefore, voltage regulation becomes important in DER systems. 
To handle this problem, the photovoltaic (PV) inverter has been used to provide reactive power, which can be converted to compensate the voltage. The reactive power is set based on the voltage measured at the point of common coupling. 
However, the voltage measurement could be missing, intermittent, unsynchronized, or unobservable, making the state estimation and voltage regulation unable to perform.
Thus, Darbali et al.\cite{darbali2021state} constructed DT to simulate DER systems based on the information of distribution system design, historical substation data, equipment configuration, and near real-time PV production data. The DT could provide pseudo-measurement  generated by real-time simulation to overcome the problem of lacking electric feeder's measurement data for DER voltage regulation.



Zhou et al. \cite{zhou2019digital} proposed a DT framework for power grid online analysis in detecting electricity failures. Traditionally, the measurement information from the RTU will be processed by SCADA and state estimation (SE) to report real-time grid change events and generate loadflow snapshots for further periodical dynamic security assessment (DSA). Statistically, the round-trip analysis takes around 10 minutes in a realistic system. 
To decrease time consumption into the order of seconds, the authors added a new data processing path through a DT-based net analysis model, which includes a Bus/Breaker model, a Node/Breaker model, and a Bus/Branch model. They are updated by SCADA and SE. The changes of DT models are considered as the grid change events and reported into a complex event processing (CEP) engine for situation awareness analysis. Once grid change events were detected, it will enable the DSA based on ML algorithms for a fast analysis.
With DT models, snapshots of power grid operation states can be created with sub-second latency without going through SE. The snapshots can be used in data-driven online analysis applications like NN-based security assessment and CEP rule evaluation that can all be completed in under 1 second. 
Thus, in a large-scale lab testing environment, a full online analysis round trip can be completed in about 10 seconds.



Focusing on the enhanced surveillance capability enabled by DT, He et al. \cite{he2018surveillance} implemented a realistic system-level DT in an industrial IoT system, Pavatar, for surveillance  and  remote  diagnosis of the ultrahigh-voltage converter station (UHVCS). Pavatar constructed DT models for real-time monitoring of UHVCS operations. It enables the sensing of compensators and cooling systems, operation environment, and human activities. Besides, it realizes data visualization based on VR technology and enables anomaly detection, root-cause diagnosis, and system error prediction with the assistant of AI analysis.


%
The development of DER and microgrids has increased the complexities of data communication and processing.
To solve this problem, a DT-based power grid architecture is promising in realizing the interaction with smart grid physical components, collecting system information, monitoring different power systems, participating in the control operations, and even detecting cyberattacks at the communication layer to ensure optimal performance of the physical grid.
Saad et al. \cite{saad2020iot} proposed the DT architecture of energy CPS. The DTs are created on the AWS cloud. The authors realized DT modeling of different power resources, interconnected microgrid system, and the communication topology.







Brosinsky et al. \cite{brosinsky2018recent} focused on using DT in the power system control centers. The authors firstly compared the SCADA system with PMU-based wide area monitoring system (WAMS) . SCADA analyzes the steady states of power grid based on the data periodically transmitted from RTU, where the data is updated in 2 to 10 seconds.
However, due to the introduction of PMU, the requirement of real-time monitoring becomes urgent. 
Usually, the data sample rate can get up to 60 per second.
So, a PMU-based WAMS becomes more capable for dynamic state analysis.
Correspondingly, a DT-based power grid control system simulating the real-time power system states can process both steady-state and dynamic-state data to enable stronger functions than SCADA. 
The DT could be a system of systems (SoS) where DTs running in different system operators could communicate with each other for collaborative operations.
It provides enhanced operational situation awareness ability.
Based on the DT and historical data, power system control centers could estimate system state even the measurement is lost or unavailable. As well, the DT power system control center could fast detect abnormal grid incidents by comparing monitored grid behaviors and simulated grid behaviors. It should provide suggestions to prevent blackout and offer an offline environment for operator training and asset optimization.

\subsection{Digital Twin as an Enabler for Enhanced Cybersecurity}


As mentioned in Subsection~\ref{sub_DT}, the object of DT simulation could be both power system units and communication network elements. Subsection~\ref{sub_DT in SG} has introduced how DT is applied in the smart grid context. It can be seen that most works utilizes DT to monitor the states of power system for security purposes. Similarly, DT could also be utilized to simulate the communication network to solve cybersecurity problems.
This section introduces the works exploiting DTs as an enabler for enhanced cybersecurity.




Becue et al. \cite{becue2018cyberfactory} introduced an ITEA project named CyberFactory\#1\cite{ITEACyberFactory}. The project aims to enhance the security and resilience of the digital factory through the collaboration of DT and cyber-range. It improves the testing capability by generating attacks on the DT model to discover potential vulnerabilities and assess the impact of attacks to support decision-making. Besides, it enhances the training capability by building cybersecurity competition with “blue team/red team” exercises on the DT models and simulating cyber-incident to help understanding its effect.

Atalay et al. \cite{atalay2020digital} firstly reviewed existing smart grid security standards and common cybersecurity threats, e.g., DDoS and APT. Then, the authors indicated the lack of security evaluation standards for the smart grid. Therefore, they proposed a DT-based security testing framework. 
The proposed evaluation framework involves i) an extensible DT for smart grid, ii) the TI database for both smart-grid-specific and common cyberattacks, iii) attack simulation tool set, and iv) A data analysis and reporting module to infer smart grid vulnerabilities and evaluate the risks.
Specifically, the DT consists three layers, i.e., physical, virtual, and decision-making layers.
The physical layer is composed of the terminal, perception, storage, computing, and networking resources in a physical power grid.
The virtual layer contains the physical and logical profiles of system components, network state, and optimization parameters.
The decision-making layer maps the physical layer to the virtual layer,optimize smart grid's operation, manage the communication topology, and generate alarms while detecting anomalous behaviors.
Based on the precise DT model, the framework provides a platform for cybersecurity test without affecting the real physical grid.



%
Networked microgirds are affected by both individual and coordinated cyberattacks, but lack of protection especially against coordinated attacks.
Saad et al. \cite{saad2020implementation} indicated that a real-time data driven model should be applicable to detect coordinated attacks and provide autonomous recovery. 
To ensure the resiliency of networked microgrids, the authors proposed an IoT-based DT framework to provide a centric oversight for the networked microgrid system and detect coordinated attacks with integrated data.
The framework is implemented on a cloud platform covering the digital replica of physical sensors, cyber controllers, and their interactions.
According to the test on a distributed control system, the authors demonstrated the effectiveness of the DT framework in mitigating FDI and (DDoS) attacks. Once the attack is discovered, observers can take corrective actions based on what-if scenarios to ensure the operation of a networked microgrid.

Bitton et al. \cite{bitton2018deriving} proposed to build cost-efficient DT as a testbed for the security assessment of specific industrial control system (ICS).
Penetration testing, as an important way for security assessment, could discover the vulnerabilities in a target network, including login services with default password, hosts with vulnerable software, improper network configuration, etc.
For ICSs, the penetration testing more focuses on the test of industrial components (e.g., PLC, sensors, HMI) and dedicated industrial protocols (e.g., Modbus, DNP3).
However, normal penetration testing operations are not practical for ICSs. Operations like vulnerability discovery and port scanning could possibly result in the break down of the target system which are apparently not feasible for ICSs.
As a result, testbeds are desired for ICS security assessment. 
Nevertheless, it is necessary to consider the trade-off between developing budget and fidelity. Thus, most of the existing testbeds are generic. To develop specific testbeds for ICSs, DT seems to be a promising way based on its simulated ICS components. 
In \cite{bitton2018deriving}, Bitton et al. proposed a way to find out the optimal fidelity of DTs for specific ICS security tests. The authors considered the budget and fidelity of DT as an optimization problem that aims to establish a DT model to executing the most important tests within a limited budget. Then, the optimal DT specification is solved with 0-1 non-linear programming.

Hadar et al. \cite{hadar2020cyber} proposed the cyber DT, which models the corporate network security situation and hacker activities. In addition to the monitoring of the computer network states, the cyber DT constructs attack graph models to realize the simulation of real-world attacks, the correlation between the network status and the attack tactics, the evaluation of the network risk value, and the acquisition and optimization of security controls’ requirements.


Taking the electric power ecosystem as a case, Salvi et al. \cite{salvi2022cyber} proposed a cyber-resilience model for critical cyber infrastructures (CCI) based on the implementation of DT.
To ensure the resilience, the model should integrate both prevention system and response strategies.
Firstly, the proposed framework contains a technical layer, a operational layer, and an ecosystem layer.
The technical layer represents the real-world CCI. While an incident occurs, detection and response processes should be triggered to identify the incident and protect the system through recovery and system hardening.
The operational layer is responsible for the coordination, communication, control, and intelligence processes.
Moreover, since CCI is usually a part of a large ecosystem, the detection of an incident should be noticed to other ecosystem units like organizations, government institutes, and law enforcement agencies. Thus, in the ecosystem layer, these units could coordinate to set standards and preventive requirements to protect their system from further affected by the threats.
Then, with the introduction of DT, the model could mimic the original CCI to provide a replicated environment with enhanced situational awareness capability for better response coordination and cyber incident management. As well, the DT provides a cyber-range platform to train CCI personnel and validate preventive control and response strategies.


Eckhart et al. \cite{eckhart2018specification} proposed a passive state replication approach to generate functional equivalent virtual entities for real-time monitoring and intrusion detection. Differently, Gehrmann et al. \cite{gehrmann2019digital} adopted state replication for active monitoring to support security analysis.
Gehrmann et al. \cite{gehrmann2019digital} focus on the architecture design of the secure DT-based industrial automation and control systems (IACS), and pays special attention to the state synchronization model. The authors pointed out that the core of the DT lies in the state replication, so it is necessary to ensure that the entire process of system state and input synchronization is accurate. 
The author proposed that the secure architecture requires synchronization security, synchronization without delay, protecting DT's external connection, access control, reliable software, isolated network for the local factory, and DT resilient to DoS.
The DT network adversary model provided by the authors indicates that, assuming DT runs in a third-party cloud with secure execution environment and data storage, then attackers may hijack, tamper, and replay the communication between physical-twin or twin-twin entities and send arbitrary requests to the DT.
Based on the above security requirements and adversary models, the authors proposed their security architecture for the DT system.
Firstly, assuming that DT runs in a secure isolated environment (e.g., a virtual machine in a secure cloud), DTs should only accept two direct external network interactions, which are synchronization and external request or response exchanging.
DTs can access to a secure clock for precise synchronization and the external interactions requires to be protected by dedicated gateway, cloud VPN, and secure communication protocols.
Besides, the states of multiple DTs should be aggregated in a common security analysis component. 
Authorized external analyzers could access all the states in the DT-based IACS. Once abnormal activities are detected, analyzers could make instant response to ensure the resilience. 
Moreover, IDS, access control, and software vulnerability discovery are also required to ensure the reliability of both network and physical entities' software. 
In the DT-based IACS architecture, DT aims to reflect the states of its physical counterpart and protect it from direct external threats. Analyzers could discover cyberattacks in the DT level and prevent them before reaching physical parts.

\section{Security Considerations of Digital Twin-based Smart Grid: Lessons Learned and Future Perspective}
\label{Section: Digital Twin Enhanced Security-Lesson Learned and Future Perspective}


The smart grid is a typical CPS. The development towards an intelligent, digital, and Internet-connected CPS also expanded its threat surface. Therefore, the cybersecurity becomes vital important. It can be enhanced from two ways. Firstly, the existing critical defense approaches need to be further improved. This can be achieved by leveraging advanced AI algorithms or solving their corresponding challenges as presented in Section~\ref{Section: Defense}. 
Secondly, novel technologies need to be applied in the smart grid context for security purposes. As mentioned in Section~\ref{DT in SG}, DT is a promising technology for enhanced cybersecurity. Thus, DT should be integrated into the security architecture design of the smart grid and develop security applications based on it.
However, each step towards a digital smart grid is also accompanied by additional security issues. Consequently, protections are indispensable for the DT itself, especially its running environment and data communication processes.

\subsection{Embedding DT into the security architecture of the smart grid}

DT has been proved as an enabler for enhanced cybersecurity. Thus, it should be embedded into the security architecture of the smart grid. 
As expected by Gehrmann et al. \cite{gehrmann2019digital}, DT should run on a secure isolated cloud environment, reflecting the states of its physical counterpart and protecting smart grid from direct external threats. 
In a DT-based security architecture, external requests would be processed in the DT so that analyzers could discover cyberattacks in the DT level and prevent them before reaching physical parts.
Besides, DT provides an uniformed interface for authorized security analysts. DTs could communicate with each other and upload the system states into an integration server. Thus, DT offers enhanced situational awareness capability to security analysts for detecting DDoS and APT.
Based on the architecture, cybersecurity applications can be further developed.
As presented in Table~\ref{tab:DT works}, DT supports data visualization based on VR. As well, it achieves the surveillance of system states and detects anomalies by comparing the real-world states with DT estimated data.
Since the system behavior simulated by DT is formulated according to the engineering design of power devices, the DT-based anomaly detection can be developed as the specification-based IDS. 
Additionally, due to the high fidelity of DT models, DT would be a good choice for high-fidelity honeypots to distract adversaries from reaching the real system.
Furthermore, DT is promising in building cyber-range for specific power systems. 
Proper fidelity of the DT should be firstly decided according to the budget. It could be in the unit level, system level, or SoS level. In the DT-based cyber-range, developers could restore system states for personnel training and penetration testing without affecting the operation of physical power systems.

\subsection{Deploying defense approaches for DT's own security}

The introduction of DT could also bring cybersecurity issues.
DT components like the DT model, data, and communication processes could become the potential targets for smart grid adversaries to deliver attack tools for gathering private information, disrupting market behavior, damaging power system operations, and causing large-scale power outages.
However, there lacks relevant work discussing the protection of DT components.
Future works should specify the protection standards of DT and required defense approaches.
For example, the DT could be a software deployed on a cloud VM. Thus, security measures like bug discovery need to be deployed to ensure the software and cloud security. 
Besides, DT requires additional communication channels to interact with physical counterparts for state synchronization or connect with external servers for receiving requests and responses or sharing TI. 
Thus, a secure synchronization clock, synchronization gateways, VPN, and encryption protocols (e.g., IPsec, TLS/DTLS) are needed to prevent eavesdropping and tampering\cite{gehrmann2019digital}.
The IDS is also needed to detect external intrusions (e.g., DoS, MITM, etc.) for protecting the security of DT communications.
For the DT data, access authorities should be strictly limited. Meanwhile, the terminal devices need to be trustworthy to ensure the credibility of data sources.
As a result, the protection of DT needs to deploy the critical defense approaches discussed in the paper for systematic passive-active protection.

\section{Conclusion}
\label{Section: Conclusion}

Cybersecurity of the smart grid have attracted the attention of academics, industries, and society. 
In this survey, we introduce the background knowledge of the smart grid. Besides, we review cyberattacks targeting the smart grid in the last decade and discuss the prevalent attack technologies. To improve the resilience, efforts should be made by either improving existing defense approaches or applying novel developed technology, like the DT, to the smart grid context. Thus, on the one hand, we review the critical defense approaches that provide passive-active protection for the smart grid by identifying smart grid assets, discovering potential software vulnerabilities, detecting intrusions, actively consuming adversaries' resources, analyzing attack features, tracing attack paths, identifying attackers, and generating TI. 
On the other hand, we review DT's existing works to demonstrate the capability of DT in enhancing smart grid cybersecurity. Based on the real-time replication of DT, it is expected to provide uniformed data interface and better situational awareness capability. Besides, it could be developed for enhanced IDS, honeypot, and cyber-range to enforce the protection of smart grid. However, the security of DT itself should also be looked after. The DT model running in a cloud environment, the stored DT data, and communication processes have to be defended to ensure the trustworthiness of DT. Future works should focus on these aspect to construct a security architecture of DT-based smart grid.

\section*{Acknowledgment}


We appreciate the support of the National Key R\&D Program of China under Grants No. 2020YFB1807500, No. 2020YFB1807504, and National Science Foundation of China Key Project under Grants No. 61831007.

\ifCLASSOPTIONcaptionsoff
  \newpage
\fi


\bibliographystyle{IEEEtran}
\bibliography{IEEEabrv,references}

%



%

\begin{IEEEbiography}[{\includegraphics[width=1in,height=1.25in,clip,keepaspectratio]{TZheng}}]{Tianming Zheng}
received the B.S. degree in the School of Automation and Electrical Engineering from the University of Science and Technology Beijing, China, and the M.S. degree in Electrical and Computer Engineering from the University of Illinois Chicago, USA. He is currently pursuing the Ph.D. degree at the School of Electronic Information and Electrical Engineering in Shanghai Jiao Tong University. His current research interests include smart grid cyber security, Internet of Things, 6G network, and artificial intelligence.
\end{IEEEbiography}

\begin{IEEEbiography}[{\includegraphics[width=1in,height=1.25in,clip,keepaspectratio]{MLiu}}]{Ming Liu} 
is a research associate at the School of Electronic Information and Electrical Engineering, Shanghai Jiao Tong University, China. His research interests include cyber threat intelligence, intrusion detection systems, and scalable data analytics.
\end{IEEEbiography}

\begin{IEEEbiography}[{\includegraphics[width=1in,height=1.25in,clip,keepaspectratio]{DPa}}]{Deepak Puthal}
is a lecturer (assistant professor) with the School of Computing, Newcastle University, United Kingdom. Before this position, he was a lecturer (2017–2019) with the University of Technology Sydney (UTS), Australia and an associate researcher (2014–2017) at Commonwealth Scientific and Industrial Research Organization (CSIRO Data61), Australia. He has a Ph.D. (2017) from the Faculty of Engineering and Information Technology, University of Technology Sydney. His research spans several areas in cyber security, blockchain, Internet of Things, and edge/fog computing and has received several recognitions and best paper awards from the IEEE.
\end{IEEEbiography}

\begin{IEEEbiography}[{\includegraphics[width=1in,height=1.25in,clip,keepaspectratio]{PYi}}]{Ping Yi}
is Associate Professor at School of Electronic Information and Electrical Engineering, Shanghai Jiao Tong University in China. He received the M.S. degree in computer science from the Tongji University, Shanghai, China. He received the Ph.D degree at the department of Computing and Information Technology, Fudan University, China. His research interests include mobile computing and artificial intelligence security. He is a member of IEEE Communications and Information Security Technical Committee.
\end{IEEEbiography}

\begin{IEEEbiography}[{\includegraphics[width=1in,height=1.25in,clip,keepaspectratio]{YWu}}]{Yue Wu}
received the B.S. degree from Dept. of Information and Electronics, Zhejiang University, Hangzhou, China in 1989, MS and Ph.D degree from Dept. of Radio Engineering, Southeast University, Nanjing, China in 1998 and 2004 respectively. He is currently a Professor with School of Electronic Information and Electrical Engineering, Shanghai Jiaotong University, Shanghai, China. His research interests include vehicular networks, wireless network security, security and trust for IoT. He is a member of IEEE and IEEE Communications and Information Security Technical Committee.
\end{IEEEbiography}

\begin{IEEEbiography}[{\includegraphics[width=1in,height=1.25in,clip,keepaspectratio]{XHe}}]{Xiangjian He} 
is currently a Professor of computer science with the School of Electrical and Data Engineering, University of Technology Sydney (UTS), and a Core Member, Global Big Data Technologies Associate Member with the AAI—Advanced Analytics Institute. As a Chief Investigator, he has received various research grants, including four national Research Grants awarded by the Australian Research Council (ARC). He is also the Director of the Computer Vision and Pattern Recognition Laboratory, Global Big Data Technologies Centre (GBDTC), UTS. He is also a Leading Researcher in several research areas, including big-learning based human behaviour recognition on a single image, image processing based on hexagonal structure, authorship identification of a document, and a documents components, such as sentences, and sections, network intrusion detection using computer vision techniques, car license plate recognition of high speed moving vehicles with changeable and complex background, and video tracking with motion blur. He has been a member with the IEEE Signal Processing Society Student Committee. He has been awarded the Internationally Registered Technology Specialist by the International Technology Institute (ITI).
\end{IEEEbiography}







\end{document}